\documentclass[acmsmall,screen,nonacm]{acmart}
\usepackage{graphicx} 
\usepackage{url}
\usepackage{svg}
\usepackage{caption}

\title{Towards Verified Code Reasoning by LLMs}
\author{Meghana Sistla}
\authornote{Work done while interning at Google.}
\affiliation{
\institution{University of Texas at Austin}
\country{USA}
}
\email{mesistla@utexas.edu}

\author{Gogul Balakrishnan}
\affiliation{
\institution{Google DeepMind}
\country{USA}
}
\email{bgogul@google.com}

\author{Pat Rondon}
\affiliation{
\institution{Google}
\country{USA}
}
\email{rondon@google.com}

\author{Jos\'{e} Cambronero}
\affiliation{
\institution{Google}
\country{USA}
}
\email{jcambronero@google.com}

\author{Michele Tufano}
\affiliation{
\institution{Google}
\country{USA}
}
\email{tufanomichele@google.com}

\author{Satish Chandra}
\authornote{Work done while at Google.}
\affiliation{
\institution{Meta Platforms}
\country{USA}
}
\email{schandra@acm.org}

\date{}

\usepackage{refs}
\usepackage{listings}

\usepackage{xcolor} 
\usepackage{caption} 
\usepackage{float}   

\definecolor{lightyellow}{rgb}{1, 1, 0.88}

\newcommand{\Souffle}{{Souffl\'{e} }}
\newcommand{\CodeSemantics}{\texttt{CodeSemantics}}
\newcommand{\AgentClaims}{\texttt{AgentClaims}}
\newcommand{\VerificationCondition}{\texttt{VerificationCondition}}
\usepackage{multirow}

\usepackage{amsmath}

\usepackage{subcaption}

\usepackage{adjustbox}
\usepackage{xspace}
\usepackage{xcolor}
\usepackage{wrapfig}
\usepackage{tabularx}
\usepackage{tablefootnote}

\usepackage{placeins}

\begin{document}

\begin{abstract}
While LLM-based agents are able to tackle a wide variety of code reasoning questions, the answers are not always correct. This prevents the agent from being useful in situations where high precision is desired: (1) helping a software engineer understand a new code base, (2) helping a software engineer during code review sessions, and (3) ensuring that the code generated by an automated code generation system meets certain requirements (e.g. fixes a bug, improves readability, implements a feature).

As a result of this lack of trustworthiness, the agent's answers need to be manually verified before they can be trusted. Manually confirming responses from a code reasoning agent requires human effort and can result in slower developer productivity, which weakens the assistance benefits of the agent. In this paper, we describe a method to automatically validate the answers provided by a code reasoning agent by verifying its reasoning steps. At a very high level, the method consists of extracting a formal representation of the agent's response and, subsequently, using formal verification and program analysis tools to verify the agent’s reasoning steps.

We applied this approach to a benchmark set of 20 uninitialized variable errors detected by sanitizers and 20 program equivalence queries. For the uninitialized variable errors, the formal verification step was able to validate the agent's reasoning on 13/20 examples, and for the program equivalence queries, the formal verification step successfully caught 6/8 incorrect judgments made by the agent.
\end{abstract}

\maketitle

\section{Introduction}
\label{Se:intro}





When writing code, whether adding a new
feature or fixing an existing bug, or when reviewing code, software engineers
must be able to carefully reason about code properties. 
Some examples of such code reasoning questions, which come up during software engineering activities, are:

\begin{itemize}
\item Does any call site depend on the order of elements in the vector returned by this method?
\item Will we avoid a copy, if we use \texttt{std::move(*example.mutable\_features())} instead of \texttt{example.features()}?
\item Can \texttt{read\_response\_settings} be null here?
\item Can \texttt{device.device\_ip()} be an invalid IP address? If so, show me a call sequence that causes an invalid ip address.
\item Can you explain why there is a use of an uninitialized value here?
\item Does this change fix the bug?
\end{itemize}

Answering these questions requires understanding the code (e.g., its design, programming language) and the question posed. Because software developers very often encounter these questions in settings such as new code bases or judging code generated by an automated (and possibly unreliable) source, there is a substantial opportunity to provide reasoning assistance.

Modern LLMs, and more recently LLM-powered agents, offer a compelling tool for assistance in such code reasoning questions. However, as is well known, LLMs (and agents that use them) have a propensity to hallucinate and may produce incorrect answers. In the domain of code reasoning, such incorrect answers can at best result in wasted developer time and at worst result in dangerous misunderstandings with the possibility of introducing serious code errors.

One way to take advantage of the code understanding abilities of LLMs, while producing high-quality
code reasoning answers is to employ formal verification over the agent's responses. In this paper, we describe how to build an LLM agent to answer code reasoning questions in such a \emph{verifiable} manner. We now describe how one might build a basic code reasoning agent, without verification, and show this can introduce subtle mistakes. We then build on this to introduce our verification approach.

\subsection{Basic Agentic Code Reasoning}
\label{Se:basic_agent}

\begin{wrapfigure}{R}{0.54\textwidth}
  \centering
  \includegraphics[width=\linewidth]{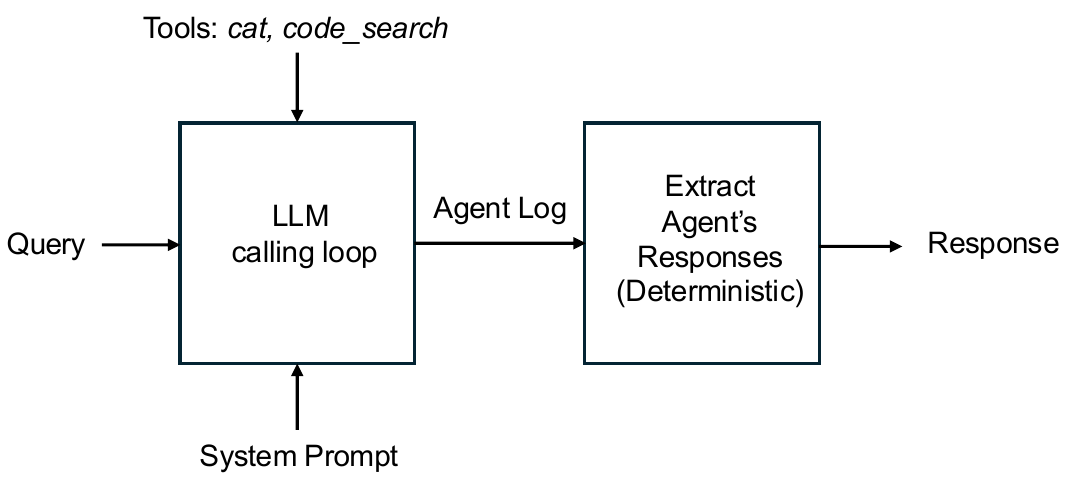}
  \caption{Basic agentic framework with an LLM function calling loop and extraction of the agent's response.}
  \label{Fi:basic_agent_architecture}
\end{wrapfigure}

\figref{basic_agent_architecture} shows the high-level architecture of the code reasoning agent. The agent is provided with tools \textit{cat} and \textit{code\_search}. The agent uses these tools to explore the code base and iteratively grab code snippets that could help answer a code reasoning question. (This style of iterative discovery of context has become standard in many applications.)  

We use a ReACT-style loop~\cite{yao2023react} to implement the agent. At every step, the model either responds with a function call request, a textual response, or both. Whenever the model’s response includes a function call request, we execute the function and add the result of the execution as a user response in the chat session. We continue these steps until either the number of allowed steps is exhausted or the model’s response contains no function call requests. Here is an example trace of what this interaction looks like:
\begin{lstlisting}[language={}]
<user> why is this null here? 
<model> user is asking the why a pointer is null. I need to find the source code first. 
<model> [function_call_request] get somefile.cc for me.
<user> [tool_output] Here is somefile.cc ...
<model> Now that I have the file, I need to understand the code. 
<model> [function_call_request] get all the uses of this variable...
<user> [tool_output] here you go...
...
<model> Here is the final response. The pointer `foo` is null because...
    
\end{lstlisting}


When the function calling loop terminates, we extract all of the model's textual responses from the agent log. The extracted sequence of messages collates the agent's thought and justifications. More details on the system prompt used in these agentic code reasoning tasks can be found in~\sectref{basic-system-prompt}.

\subsection{But can you trust it?}

When the agent was run on a set of code-reasoning questions from CRQBench~\cite{dinella2024crqbench} (a benchmark of 100+ code-reasoning questions), the agent was able to tackle a wide variety of questions, but the answers were not always correct. For example, the agent may say two fragment of code  mean the same where in reality they do not.  (Several examples appear in Section~\ref{Se:errors}.)
This prevents the agent from being useful 
because
the answers need to be manually verified before they can be trusted. Verifying the responses from the code reasoning agent manually involves human effort and can lead to slower developer productivity, which is undesirable.


The main contribution of the paper is a method to automatically validate the answers provided by the agent by verifying its reasoning steps.
At a very high level, the method consists of converting the textual output of the explanation from the agent into a formal representation and, subsequently, using formal verification and program analysis tools to verify that the agent’s reasoning steps do not contain any errors.

We applied this approach to a benchmark set of 20 uninitialized variable errors detected by memory sanitizers and
20 program equivalence queries. For the uninitialized variable errors, the formal verification step was able
to validate the agent’s reasoning on 13/20 examples, and for the program equivalence queries, the formal
verification step successfully caught 6/8 incorrect judgments made by the agent.

The paper is organized into the following sections:
\begin{itemize}
    \item \sectref{errors} delves into the kinds of errors we notice in LLM outputs when attempting to answer code reasoning questions. 
    \item The general framework used for verifying LLM reasoning is discussed in~\sectref{verified}.
    \item \sectref{uninit-vars} shows a demonstration of our approach to a simpler problem of memory-sanitizer bugs.
    \item \sectref{program-equivalence} discusses our work in more detail for the problem of program equivalence.
    \item \sectref{future} discusses future work,~\sectref{related} discusses related work, and~\sectref{conclusion} concludes.
\end{itemize}

\section{Errors in LLM Reasoning}
\label{Se:errors}



In this section, we will discuss the type of errors shown in LLM reasoning in the context of answering code-reasoning questions.
A manual evaluation of the agent’s responses shows that the agent can make errors in its reasoning steps.  Broadly, the errors fall into one of the following categories:
(a) the agent makes incorrect assumptions about code semantics;
(b) the agent fails to consider all aspects of the problem necessary to justify the final response; or 
(c) the agent makes unwarranted assumptions about library behavior.

\subsection{Incorrect Assumptions About Code Semantics}

Consider the code shown in~\figref{arraybounds1}, where \texttt{str[i]} has an out-of-bounds access if \texttt{start >= len}.

\begin{figure}[t!]
    \centering
    \noindent 
    \begin{lstlisting}
absl::StatusOr<unsigned> find(unsigned start, char *str, char goal) {
  unsigned len = strlen(str);
  for (unsigned i = start; str[i] != 0 && i < len; i++) {
    char *tmp = str;
    if (tmp[i] == goal) return i;
  }
  return absl::NotFoundError("goal not found");
}
    \end{lstlisting}
    \vspace{-2ex}
    \caption{An array-bounds example where the code reasoning question (asked to the agent) is asked ``Is \texttt{str[i]} access safe?''; the correct answer is \textit{unsafe}.}
    \label{Fi:arraybounds1}
\end{figure}

The agent is asked the following question: “Assume that str is (1) not null, and (2) properly null-terminated. Is the access \texttt{str[i]} safe?”
In several runs, Gemini (1.5 pro~\cite{team2024gemini}/2.0-flash~\cite{comanici2025gemini}) makes an incorrect assumption that the condition \texttt{i < len} strictly dominates the access \texttt{str[i]} and reports that there are no memory errors in the program, as shown in~\figref{arrayboundswrong}. 

\begin{figure}[t!]
    \centering
    \noindent 
    \begin{lstlisting}[language={}]
// WRONG: Makes incorrect assumption that i < len strictly dominates str[i] != 0.
The loop condition is `str[i] != 0 && i < len`.
- `str[i] != 0` checks if the character at index `i` is not the null terminator.
- `i < len` checks if `i` is within the bounds of the string, where `len` is the length of the string (excluding the null terminator).

Since `str` is assumed to be null-terminated, the loop will terminate when `str[i]` is 0. Also, the condition `i < len` ensures that `i` does not exceed the length of the string. Therefore, the access `str[i]` is safe because it will always be within the bounds of the string or at the null terminator.

Final Answer: The access `str[i]` is safe.
    \end{lstlisting}
    \vspace{-2ex}
\caption{LLM explanation for code in~\figref{arraybounds1}. The agent makes false assumptions about code semantics and incorrectly answers the code reasoning question.}
\label{Fi:arrayboundswrong}
\end{figure}

However, in other invocations, the agent produces a correct response. \figref{arrayboundscorrect} shows an abridged example where the LLM produces the correct answer.

\begin{figure}[t!]
    \centering
    \noindent 
    \begin{lstlisting}[language={}]
// CORRECT: Considers various scenarios, incl. str[i] being an out-of-bound access.
The loop condition is `str[i] != 0 && i < len`.
Lets analyze the safety of `str[i]`:

. . .

If `i` is equal to `len`, then `i < len` is false, and the loop terminates. However, before that, `str[i] != 0` is evaluated. Since `len` is the length of the string (excluding the null terminator), `str[len]` is the null terminator, so `str[len]` is a valid access and `str[len] == 0`.
. . .

Therefore, the access `str[i]` is safe as long as `start` is a valid starting index.
    \end{lstlisting}
    \vspace{-2ex}
\caption{Correct Explanation for code in~\figref{arraybounds1}, the agent considers various scenarios, doesn't make incorrect assumptions, and correctly answers about the code reasoning question.}
\label{Fi:arrayboundscorrect}
\end{figure}

\subsection{Failing To Consider All Aspects of the Problem}

\begin{figure}[t!]
    \centering
    \noindent 
    \begin{lstlisting}
char foo(char * x, int len, int index) {  
  if (index < len) {
    return x[index];
  } else {
     return 0;
  }
}
    \end{lstlisting}
    \vspace{-2ex}
\caption{Another array-bounds example where the agent is asked the code reasoning question - ``Is the access x[index] safe?'', and the answer is that the access is potentially \textit{unsafe}.}
\label{Fi:arraybounds2}
\end{figure}

Another common failure mode is that the agent may not consider all the aspects necessary to justify the final response. Consider the code snippet shown in~\figref{arraybounds2}. The agent is asked the following question: “len is the size of the buffer x.  Is the access x[index] safe?”. The access is potentially unsafe for the following reasons:
\begin{itemize}
\item x could be a nullptr.

\item Either len or index could be negative and cause an out-of-bounds memory access. 
\end{itemize}

The agent should take into account all of the above aspects before answering the question. However, the agent does not always do so as can be seen in~\figref{arrayboundswrong2}.

\begin{figure}[t!]
    \centering
    \noindent 
    \begin{lstlisting}[language={}]
// WRONG: Did not consider all aspects of the problem
The code checks if index is less than len before accessing x[index]. If index is within the bounds of the buffer (i.e., index < len), then the code returns the character at that index. Otherwise, if index is out of bounds, the code returns 0, preventing an out-of-bounds access.
    \end{lstlisting}
    \vspace{-2ex}
\caption{Incorrect LLM explanation for code in~\figref{arraybounds2}; the explanation is incorrect because the agent did not consider all possible scenarios of why \texttt{x[index]} could be unsafe -- (1) x is a null-pointer, (2) \texttt{index} or \texttt{len} could be negative.}
\label{Fi:arrayboundswrong2}
\end{figure}

There are other instances where the model answers correctly, shown in~\figref{arrayboundsright2}.

\begin{figure}[t!]
    \centering
    \noindent 
    \begin{lstlisting}[language={}]
// CORRECT: considers all aspects of the problem.
There is no immediately obvious problem with this code. It checks if the index is within the bounds of the array and returns the element at that index if it is, otherwise it returns 0. However, there are some potential issues to consider:

1.  **Null pointer `x`:** The code doesnt check if `x` is a null pointer. If `x` is null, then `x[index]` will cause a segmentation fault.
2.  **`len` is non-positive:** If `len` is zero or negative, and `index` is non-negative, then the condition `index < len` will evaluate to false if `len` is zero, and it will evaluate to true if `len` is negative. If `len` is negative, then `x[index]` will access memory out of bounds.
3.  **`index` is negative:** If `index` is negative, then `x[index]` will access memory out of bounds.

    \end{lstlisting}
\caption{Correct LLM Explanation for code in~\figref{arraybounds2}}
\label{Fi:arrayboundsright2}
\end{figure}

\subsection{Making unwarranted assumptions about library behavior}
\label{Se:errors-library}
Another scenario is where the agent makes unwarranted assumptions about library code behaviour, which can lead to incorrect
responses from the agent.


Consider the two code snippets shown in~\figref{program-equivalence-example} that are similar except for how the condition on Line 82 is evaluated.
In Code Snippet 1, the value of $\tt has\_value$ is obtained by calling $\tt session\_input.has\_value()$ on Line 80;
whereas in Code Snippet 2, the value of $\tt is\_empty$ is obtained by calling $\tt IsEmpty(session\_input)$ on Line 80.

\begin{figure}[t!]
    \centering
    \noindent 
\begin{subfigure}[t]{0.48\linewidth}
    \begin{minipage}[t]{\linewidth}
    \begin{lstlisting}
// main.cpp
...
80: bool has_value =
81:       session_input.has_value();
82: if (has_value) {
83:   status.check = 0;
84:   status.type =
85:     GetType(session_input.value());
86: }
...
    \end{lstlisting}
\end{minipage}
\caption{Code Snippet 1 using \texttt{has\_value} method.}
\label{Fi:program-equivalence-example-a}
\end{subfigure}
\hfill 
\begin{subfigure}[t]{0.48\linewidth}
    \begin{minipage}[t]{\linewidth}
    \begin{lstlisting}
// main.cpp
...
80: bool is_empty =
81:       IsEmpty(session_input);
82: if (!is_empty) {
83:   status.check = 0;
84:   status.type =
85:     GetType(session_input.value());
86: }
...
    \end{lstlisting}
\end{minipage}
\caption{Code Snippet 2 using \texttt{IsEmpty} method.}
\label{Fi:program-equivalence-example-b}
\end{subfigure}

\caption{Two \textit{not-equivalent} code snippets are shown where the condition at Line 82 is obtained by checking for ${\tt has\_value()}$ 
in Code 1 and ${\tt IsEmpty()}$ in Code 2. The agent is asked ``Are the code snippets equivalent?''.}
\label{Fi:program-equivalence-example}
\end{figure}

On querying the agent about the equivalence of the two code snippets, the agent incorrectly assumes the behavior of the function calls $\tt has\_value()$ and $\tt IsEmpty()$, and falsely concludes the programs

Given the two code snippets, the agent is asked the following question: "Are the two code snippets equivalent? Explain your reasoning.". The two code snippets are not equivalent as the calls $\tt has\_value()$ and $\tt IsEmpty()$ could have different behavior. The agent's output is as shown in~\figref{explanation1}. The agent clearly hallucinates and incorrectly concludes that the programs are equivalent by justfying its claims with 
\texttt{"Based on common C++ programming patterns and standard library conventions,..."}.
The agent is aggressive in its assumptions, which leads to an incorrect response.

\begin{figure}[t!]
    \centering
\begin{minipage}[t!]{0.99\textwidth}
    \begin{lstlisting}[label={lst:explanation}, language={}]
The two code snippets are semantically equivalent. Here is a step-by-step analysis of the reasoning.
...
### Equivalence Justification
- The condition to execute the inner block in the **original code** is `session_input.has_value() == true`.
- The condition to execute the inner block in the **modified code** is `IsEmpty(session_input) == false`.

Based on common C++ programming patterns and standard library conventions, a `has_value()` member function and an `IsEmpty()` free function for the same object are designed to be logical opposites. That is, an object ""has a value"" if and only if it is ""not empty"".

Therefore, the boolean expression `session_input.has_value()` is equivalent to `!IsEmpty(session_input)`.
...
    \end{lstlisting}
\end{minipage}
    \caption{Incorrect LLM explanation for a program equivalence query for programs in~\figref{program-equivalence-example}; the agent makes incorrect assumptions about the library functions \texttt{has\_value} and \texttt{IsEmpty}.}
    \label{Fi:explanation1}
\end{figure}

Even after providing the agent with the definitions for 
$\tt has\_value()$ and $\tt IsEmpty()$; the agent continues to
hallucinate and incorrectly reason about the program equivalence.
More details can be found in~\sectref{program-equivalence-example-elaborate-section}.

\section{Verifying LLM reasoning}
\label{Se:verified}

Given a textual explanation of the reasoning steps produced by an agent, how can we trust that its findings are not hallucinations? Can we avoid the issues pointed out in the prior section? 


One of the common approaches that has emerged is the use of another LLM as a judge~\cite{zhou2024mitigating, spiliopoulou2025play} to assess the assumptions. In this approach, an LLM is provided with some criteria for correctness and asked to judge if the agent’s explanation satisfies the given criteria. This approach is not foolproof either and is still susceptible to the LLM judge hallucinating.

In this paper, we propose an alternative approach, where we extract a formal representation of the agent’s reasoning steps and use formal verification tools to verify the agent’s reasoning steps. A reader will wonder why not just
use formal verification tools to answer the original code reasoning questions to begin with?  There are several advantages to using LLMs to answer code reasoning questions:  they are very good at slicing away the non-essential details, and usually don’t get bogged down by the inherent conservativeness of static analysis.  They are quick and work directly on source code as given; they do not require elaborate set up of build files etc.  Thus, the approach we take is to validate an LLM's own reasoning steps--given in prose--that lead to an answer, rather than to attempt to answer the original code reasoning question using formal techniques.  (This is reminiscent of modern LLMs sometimes calling out to, say, a calculator, to discharge a arithmetic step; except that we do this \textit{post facto}, rather than during inference.)

\subsection{\CodeSemantics, \AgentClaims, and \VerificationCondition}

To extract and verify the formal representation of the agent's reasoning steps, we introduce three terms:
\begin{itemize}
    \item \CodeSemantics: Formal representation of the ground truth semantics of the underlying source code.
    \item \AgentClaims: Formal representation of the agent's response.
    \item \VerificationCondition: Formal representation of the properties that need to be verified in the agent's response.
\end{itemize}

The agent’s explanation is correct if the following holds:

\begin{equation}
\label{Eq:condition-1}
    \CodeSemantics \Rightarrow \AgentClaims    
\end{equation}
\vspace{-2ex}
\begin{equation}
\label{Eq:condition-2}
    \AgentClaims \Rightarrow \VerificationCondition
\end{equation}

\paragraph{Agent Claims} Given the agent's output (including explanation) for the code-reasoning question, the code base, and a chosen formal representation method, we extract the agent's claims in a formal representation. In this work, we use formal predicates as the formal representation method. More on this representation used for capturing agent's claims formally is discussed in~\sectref{formal-representation}.
The extraction method is specific to the code-reasoning question, but at a high level, the formal predicates and the pointers to the codebase are provided to an LLM-based agent to extract the agent's claims for the specific code-reasoning task. The formal predicates for each task and how the agent is prompted are discussed in the specific sections in~\sectrefs{msan-agent-claims}{equiv-agent-claims}.

\paragraph{Code Semantics} We can use compiler or static analysis tools to generate the code semantics. There are several choices for the compiler or static analysis tool. CodeQL~\cite{de2007keynote} is one such tool that can be used to faithfully capture the underlying code semantics and provide an SQL-like query language to query properties of code. However, although important, in this paper we do not discuss capturing code semantics and therefore do not verify Eq.~\eqref{Eq:condition-1}, leaving this to future work.

\paragraph{Verification Condition}
Automatically inferring the verification condition for a code reasoning question is a separate problem in itself.
For certain categories of questions, such as memory errors, the verification condition is fairly straightforward. However, it is not a straightforward task for open-ended questions.
Fortunately, many code reasoning questions are already in the form of a verification condition. Therefore, we can check for it directly. For other questions, the verification condition may not be available directly and needs to be inferred. For example, consider the problem of checking the equivalence of two program snippets; the verification condition would consist of the syntactic and semantic properties that should remain consistent between the two program snippets. The set of syntactic and semantic properties would depend on the code snippets themselves and, therefore, would need to be inferred using other means. 
In this paper, we will assume that the verification condition is defined and given by the user. 
We will discuss the task-specific verification conditions in~\sectrefs{msan-verification-condition}{equiv-verification-condition}.

\subsection{Background: Datalog Predicates for Formal Representation}
\label{Se:formal-representation}
We will demonstrate how a formal representation can be used to detect common errors in the agent’s reasoning steps. 
In this work, we use \Souffle’s~\cite{scholz2016fast} datalog program as the formal representation. \Souffle is a datalog-inspired logic programming language used specifically for program analysis and verification. Datalog programs~\cite{green2013datalog} have been widely used in the field of formal verification and are quite suitable to formalize the agent’s reasoning steps. Datalog consists of a set of facts (in terms of first-order predicates) and rules. 

For our use cases, Datalog programs can be used to capture agent’s reasoning as follows:

\begin{itemize}

    \item Datalog facts can be used to represent the claims made by the agent. 
    \item Datalog rules can be used to capture the inferences made by the agent’s output based on other facts.

\end{itemize}

\begin{wrapfigure}{R}{0.54\textwidth}
    \centering
            \begin{lstlisting}
            1. int a = 0;
            2. int d = 2;
            3. output(d);
            4. int c = a;
            5. output(c);
            \end{lstlisting}
    \caption{A non-branching program with only \texttt{int} datatype and one function \texttt{output} that takes a single integer variable as input. Every variable is either defined as a constant integer value or is assigned a value of another variable using a ``copy'' assignment. The goal is to check if there exists an instance of \texttt{output} function call with a non-zero input. The program has non-zero input as variable \texttt{d} to function \texttt{output} on Line 3.}
    \label{Fi:datalog-example}
    \vspace{2ex}
\end{wrapfigure}

\paragraph{Example} Consider the program in~\figref{datalog-example}.
The program is non-branching with only \texttt{int} datatype and one function \texttt{output} that takes a single integer variable as input. Every variable is either defined as a constant integer value or is assigned a value of another variable using a ``copy'' assignment (as on Line 4). The condition is that the \texttt{output} function should only have variables with value $0$ as the input.






\begin{figure}[t!]
    \centering
    \noindent 
    \begin{lstlisting}[language={}]
// Formal Predicate Signatures
1. .decl defZero(x: symbol, l: number)
2. .decl defNonZero(x: symbol, l: number)
3. .decl copy(x: symbol, y: symbol, l: number)
4. .decl outputFn(x: symbol, l: number)

5. .decl nonZeroInputToOutputFn(x: symbol)
6. .decl isUnsafe()

// Rules/ Inferences
7. defZero(x, l) :-
8.     copy(x, y, l), defZero(y, l1), l1 < l.

9. defNonZero(x, l) :-
10.     copy(x, y, l), defNonZero(y, l1), l1 < l.

11. nonZeroInputToOutputFn(x) :-
12.     outputFn(x, l), l1 < l, defNonZero(x, l1).

// Verification Condition
13. isUnsafe() :-
14.     outputFn(x, _), nonZeroInputToOutputFn(x).

// Facts (Predicate instances) -- AgentClaims
15. defZero("a", 1).
16. defNonZero("d", 2).
17. outputFn("d", 3).
18. copy("c", "a", 4).
19. outputFn("c", 5).

    \end{lstlisting}
    \caption{A datalog program for the program in~\figref{datalog-example}. Note that the line numbers are only for ease of reference in text.}
    \label{Fi:datalog-example-predicates}
\end{figure}

\figref{datalog-example-predicates} illustrates a datalog program that captures the required facts and the verification condition for the program in~\figref{datalog-example}. Note that the line numbers are only for easy reference. Lines 1-4 are the formal predicate signatures used to capture if a variable is assigned a zero or non-zero value, if the variable is defined using a ``copy'' assignment, and to capture \texttt{output} function call with a particular argument.

\texttt{AgentClaims} would be the facts or predicate instances on Lines 15-19 for the specific program~\figref{datalog-example}. The verification is to check if \textit{all} inputs to \texttt{output} are zero; however, we can frame this as a \textit{safety} problem and check for the negation. Therefore, the \texttt{VerificationCondition} (\texttt{isUnsafe} on Line 6) is to check if there exists an instance of \texttt{output} with a non-zero input. To accomplish this, we can introduce rules to arrive at this conclusion, as shown on Lines 7-14.

Note that we did not check for \texttt{CodeSemantics} in~\figref{datalog-example-predicates}. However, for this example, Condition~\ref{Eq:condition-1} would check if the facts made by the agent from Lines 15-19 are indeed true and if they truly capture the code semantics.

\section{Modeling Uninitialized Variables Errors}
\label{Se:uninit-vars}


In this section, we illustrate our approach of verifying agent claims against a verification condition for a simpler problem of memory-sanitizer (MSAN) bugs. Given a stack trace for the use-of-uninitialized-value bugs, we ask the code reasoning agent to explain the root cause and subsequently verify the agent’s response using our approach. 
In the case of MSAN bugs, we have the dynamic trace of execution as well as the fact that the use of an uninitialized value did happen, which is not typical. However, we show how our approach can be used in this context and how we can validate the agent's reponse using program analysis and formal verification tools.



We will first discuss the verification condition (\sectref{msan-verification-condition}) for this problem of memory-sanitizer bugs, and later describe how to extract \AgentClaims (\sectref{msan-agent-claims}).

\subsection{Verification Condition}
\label{Se:msan-verification-condition}
The verification condition for MSAN bugs can be defined as follows:
\begin{equation}
    \exists x\ \exists y\ \exists x_l\ \exists y_l\ uninitialized(x, x_l) \land flow(x, x_l, y_l) \land use(y, y_l)
\end{equation}

where, $uninitialized(x, x_l)$ captures that expression $x$ is uninitialized at source location $x_l$, $flow(x, x_l, y, y_l)$ captures the fact there is a data flow from variable $x$ at $x_l$ to $y$ at $y_l$, and $use(y, y_l)$ captures the fact that $y$ is used at $y_l$. 
More on these predicates is discussed in~\sectref{formal-rep-msan}.






If we can verify that the agent proved the verification condition, we can conclude that the agent’s reasoning about the root cause for the bug is correct.

\subsection{Extracting \AgentClaims}
\label{Se:msan-agent-claims}
In this section, we will describe how to extract a formal representation of the agent's reasoning steps. To extract \texttt{AgentClaims} using formal predicates, we can provide
a vocabulary set of formal (first-order) predicates, and ask the agent to use this vocabulary to generate the formal representation of the output for the code reasoning query.

\begin{figure}[t!]
  \centering
  \includegraphics[width=0.8\textwidth]{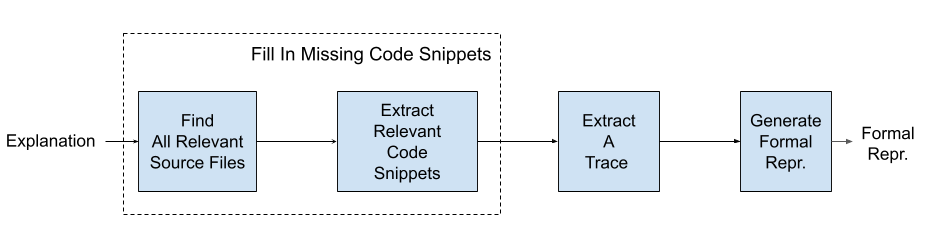}
  \caption{Generating a formal representation for MSAN bugs from the agent's explanation.}
  \label{Fi:msan_formalization_architecture}
\end{figure}

\figref{msan_formalization_architecture} shows the steps involved in generating a formal representation. The steps for extracting the formal representation for the current MSAN bugs include: (1) Fill in missing code snippets, (2) Trace Extraction, and (3) Generate Formal Representation. We first describe the structure of the formal representation and its generation and later briefly touch upon the other steps.

\subsubsection{Formal Representation}
\label{Se:formal-rep-msan}
\begin{table}[t!]
    \centering
    \resizebox{0.99\linewidth}{!}{
    \begin{tabular}{|l|l|}
        \hline
         \textbf{Predicate Set} & \textbf{Semantic Meaning} \\
         \hline
         $\tt uses(x : str, f : str, l : num)$ &  Expression \texttt{x} is used at line \texttt{l} in file \texttt{f}.\\
         \hline
         $\tt uninitialized(x : str, f : str, l : num)$ & Expression $\tt x$ is uninitialized at line $\tt l$ in file $\tt f$.\\
         \hline
         $\tt hasInitializer(x: str, m: str)$ & Variable $\tt f$ has an initializer in $\tt m$.\\
         \hline
         $\tt hasMemberInitializer(x: str, m: str)$ & Constructor $\tt m$ has an initializer for member variable $\tt x$.\\
         \hline
         $\tt allocated(x: str, f: str, l: num)$ & Memory referred to by $\tt x$ is allocated at line $\tt l$ in file $\tt f$.\\
         \hline
         $\tt declared(x: str, f:str, l:num)$ & Variable $\tt x$ is declared at line $\tt l$ in file $\tt f$.\\
         \hline
         $\tt flow(x: str, f_1: str, l_1: num, $ & The value of $\tt x$ defined at line $\tt l_1$ in file $\tt f_1$ \\
         \quad $\tt y: str, f_2: str, l_2:num)$ & flows to $\tt y$ at line $\tt l_2$ in file $\tt f_2$.\\
         \hline
         $\tt memoryError(x: str, error\_type: str,$ & There is a $\tt error\_type$ kind of memory error \\
         \quad $\tt f: str, l: num)$ & involving $\tt x$ at line $\tt l$ in file $\tt f$.\\
         \hline
    \end{tabular}
    }
    \caption{Vocabulary of formal predicates (signatures) for MSAN queries}
    \label{Ta:predicate-set-msan}
\end{table}

Before detailing the formal representation for MSAN bugs, we define the notion of a code trace.


\begin{definition}
    A code trace (or trace) is a sequence of program states or executed statements recorded during a run.
    It represents the concrete path a program takes under a specific input or execution scenario.
    In our context, a trace consists a sequence of program points, with associated code snippets, highlighting the data and control flow between these program points.
\end{definition}

We generate a formal representation of the trace by prompting the LLM-based agent with a system instruction that contains the vocabulary of formal predicate signatures (see~\sectref{system-prompts-msan} for the system prompt).~\tableref{predicate-set-msan} shows the vocabulary of predicate signatures used to represent $\tt AgentClaims$ formally.~\figref {example-trace-formal-representation} shows an example of the formal representation of a trace using the predicate signature set shown in~\tableref{predicate-set-msan}.

\begin{figure}[t!]
    \centering
\begin{lstlisting}
allocated("data_", "audio/base/audio_buffer.cc", 375).
flow("audio_buffer","client/testing/chia_client_test.cc", 28, "audio_buffer", "internal/client/client.cc", 527).
uses("audio_buffer", "internal/client/client.cc", 527).
flow("audio_buffer", "internal/client/client.cc", 527, "audio_buffer", "internal/client/client.cc", 613).
uses("audio_buffer", "internal/client/client.cc", 613).
uses("sample_ptr", "internal/client/encoders/audio_encoder.cc", 218).
flow("data_", "audio/base/audio_buffer.cc", 375, "sample_ptr", "internal/client/encoders/audio_encoder.cc", 218).
uses("buffer", "utils/encoders/stream_encoder.c", 2538).
flow("data_", "audio/base/audio_buffer.cc", 375, "buffer", "utils/encoders//stream_encoder.c", 2538).
uninitialized("data_", "audio/base/audio_buffer.cc", 375).
memoryError("data_", "uninitialized_data", "utils/encoders//stream_encoder.c", 2538).
\end{lstlisting}
\caption{Example of formal representation of a trace using the predicate set shown in~\tableref{predicate-set-msan} for MSAN tasks.}
    \label{Fi:example-trace-formal-representation}
\end{figure}

Next, we describe the process of extracting the trace from the agent’s explanation.
This is particularly useful when prompting the agent to generate a formal representation directly from the explanation is not fruitful
(due to the model size, complexity of the problem, etc.). In such cases, adding a separate post-processing step to extract a trace and thereby \texttt{AgentClaims} is useful.

\paragraph{Fill in Missing Code Snippets}
We will now use the agent's explanation, that also contains interleaved relevant code snippets, for the code-reasoning query to extract \AgentClaims. The agent is then asked to use these relevant, probably partial, code snippets in the explanation and the code base to find all the relevant source files and extract any missing code snippets, which will then be used to extract a trace.

\paragraph{Trace Extraction}
The next step to generating a formal representation is making another call to the LLM to extract a code ``trace'' that backs up the agent’s reasoning. In this LLM call, we provide the following inputs: (1) all the source code retrieved in the previous step, (2) the final summary from the reasoning agent. The system instruction for trace extraction can be found in~\sectref{system-prompts-msan}.



\subsection{Results}
\label{Se:results-msan}

We randomly selected 20 MSAN bugs that were filed  on internal projects for this evaluation. Unless explicitly stated, we ran the experiments with temperature(t) = 0.2 and $\text{top\_p}$ = 0.95 configuration. First, we ran the code reasoning agent (with Gemini 2.0 Flash model) in \sectref{basic_agent} five times and manually classified the reasoning as valid or invalid.  Out of the 100 runs (20 bugs $\times$ 5 runs per bug), 88 runs had a valid explanation based on human judgment.



\begin{figure}[htbp] 
    \centering
    \includegraphics[width=.6\linewidth]{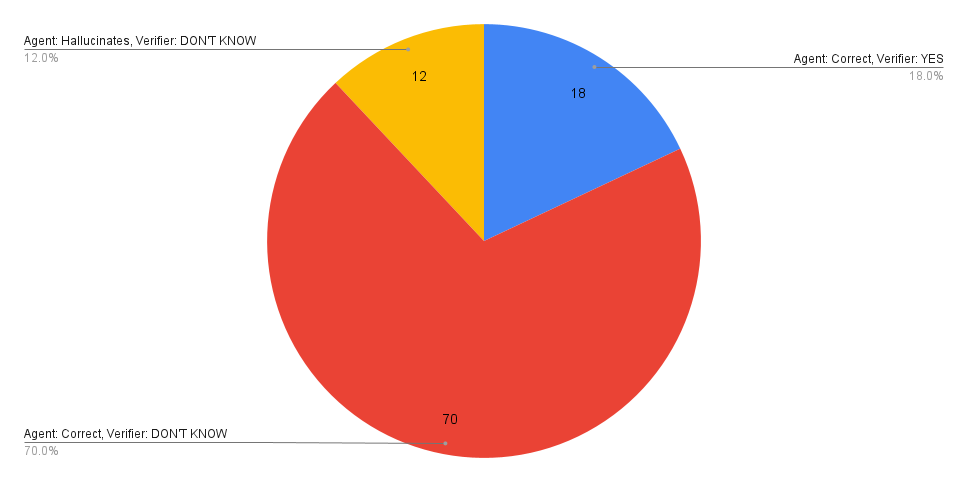}
    \caption{ \label{Fi:msan_flash2.0_results_1_iter}Results of formal verification to validate agent's explanation of the use of an uninitialized variable error, where the formalizer was run for {\bf one iteration}. If verifier says ``YES'', then it is indeed verified but if the output is ``DON'T KNOW'', then we are unsure of the agent's output.}
\end{figure}

Next, we passed the explanation of the agent’s reasoning steps to the formal verifier to check the validity of the explanation. \figref{msan_flash2.0_results_1_iter} shows the results of the formal verification check. The 100 different runs were classified in one of the following categories:
\begin{itemize}
\item {\bf Agent: Correct, Verifier: YES}: The cases where the agent was correct as per human judgment and the verifier is able to verify that the explanation is correct.
\item {\bf Agent: Correct, Verifier: DON'T KNOW}: The cases where the agent was correct as per human judgment, but the verifier is unsure of the agent's output.
\item {\bf Agent: Hallucinates, Verifier: DON'T KNOW}: The cases where the agent hallucinates as per human judgment and the verifier is unsure of the agent's output.
\end{itemize}



Note that the verifier has perfect precision and low recall, because the if verifier says ``YES'', then it is indeed verified but if the output is ``DON'T KNOW'', then we are unsure of the agent's output. As we can observe, the verification step is only able to prove the validity of 18 out of 88 valid explanations (blue). Manual investigation of the results showed that the problem was due to inconsistent outputs from the model during the formalization steps. For example, the model was not consistently generating all the datalog facts even though the textual explanations and the extracted traces contain the relevant claims. 

\begin{figure}[htbp] 
    \centering
    \includegraphics[width=.6\linewidth]{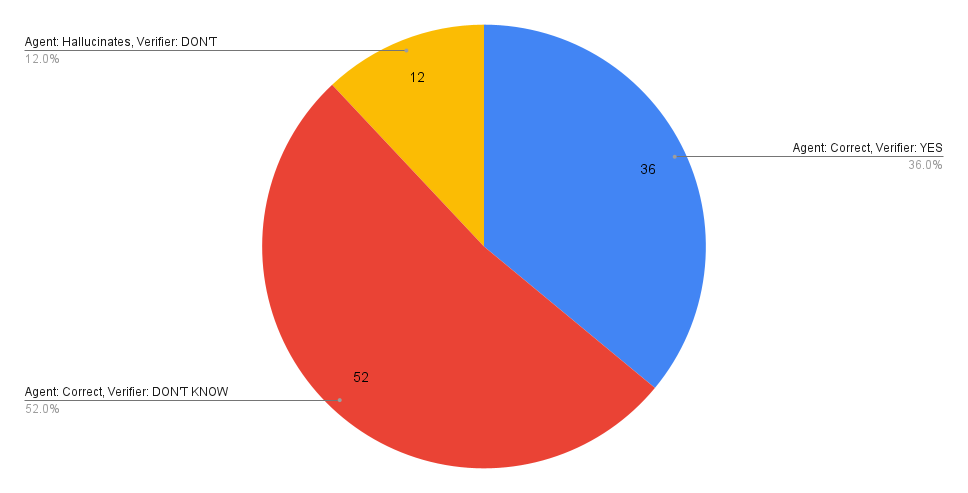}
    \caption{ \label{Fi:msan_flash2.0_results_5_iter}Results of formal verification to validate agent's explanation of the use of an uninitialized variable error, where the formalizer was run for {\bf 5 iterations}. If verifier says ``YES'', then it is indeed verified but if the output is ``DON'T KNOW'', then we are unsure of the agent's output.}
\end{figure}

One way to mitigate the problems is to run the formula (predicate instances) extraction step multiple times and combine the outputs.  The approach is sound for all practical purposes because we are only adding missing facts with each subsequent run. Any inconsistencies that are introduced in subsequent iterations will be caught by the formal checks. By running the formula extraction step five times, we were able to improve the results. In particular, the total number of verified runs went from 18/88 to 36/88 as shown in \figref{msan_flash2.0_results_5_iter} (blue).

 Among the 36 validated runs of the agent, there was at least one validated run corresponding 13 out of 20 MSAN bugs.  In other words, we were able to prove the validity of the explanation for 13 out of the 20 MSAN  bugs we considered. This could be improved further by making the model generate consistent outputs either by tweaking the prompts further or by doing additional training for style. We will defer those to future work.

We also ran the experiments with a newer, stronger model like Gemini 2.5 Pro~\cite{comanici2025gemini}. More details can be found in~\sectref{msan-eval-appendix}.

\section{Program Equivalence}
\label{Se:program-equivalence}

In this section, we show how we can verify an agent's explanations for program equivalence queries using the same framework. We first introduce the verification condition for this task and then describe how to extract agent claims, with a particular focus on what predicates are needed to express claims for this domain. Finally, we present results for equivalence queries from two benchmarks.

\subsection{Verification Condition}
\label{Se:equiv-verification-condition}

The verification condition for the problem of program equivalence is -- \emph{two programs $p_1$ and $p_2$ are considered equivalent, if for the same initial state of $p_1$ and $p_2$,
the final state of both programs (for all variables reaching the exit point) -- assuming they terminate normally -- are equivalent.}



\begin{equation}
\begin{split}
    areProgramsEquivalent(p_1, p_2) &: (\forall x : var \in p_1, watchVar(x) \\
    & \rightarrow \exists y : var \in p_2, watchVar(y) \wedge equiv(x,y)\\
                                    & \wedge (\forall y : var \in p_2, watchVar(y) \\ 
                                    & \rightarrow \exists x : var \in p_1, watchVar(x) \wedge equiv(x,y)\\
\end{split}
\end{equation}

where, $x$ and $y$ are variables in $p_1$ and $p_2$, $watchVar(x)$ indicates the variables on which equivalence is evaluated at the
exit point of the programs, and $equiv(x, y)$ is a rule that checks if the variables $x$ and $y$ are semantically equivalent. Every variable that is reachable (i.e., a variable that is alive) in $p_1$ at the exit point has an equivalent variable at the exit point in $p_2$.

Note that we use a stronger notion of equivalence than observational equivalence.
\textit{Observational Equivalence of two programs is defined as follows: Two programs $p_1$ and $p_2$ are 
equivalent (under observational equivalence) if no larger program can distinguish between the
two programs, i.e., there is no difference in terms of the observable output of the two programs~\cite{morris_lambda-calculus_1969}.} In contrast, our definition of equivalence is stronger as we check if all the variables captured using $watchVar$ are equivalent.

\subsection{Extracting \texttt{AgentClaims}}
\label{Se:equiv-agent-claims}

\begin{figure}[t!]
    \centering
    \includegraphics[width=\linewidth]{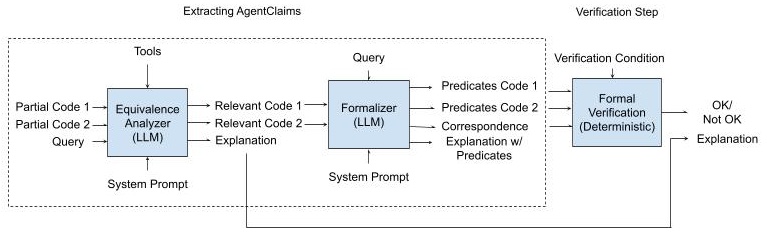}
    \caption{Equivalence queries pipeline: Analyzer $\rightarrow$ Formalizer $\rightarrow$ Verifier.}
    \label{Fi:equivalence-pipeline}
\end{figure}

We implement a two-step approach to capturing \AgentClaims~ for equivalence queries. First, we apply an equivalence analyzer, which is an LLM-based agent that provides an initial answer to the equivalence question. Second, we apply a formalizer that extracts formal statements from the equivalence analyzer's response. We now describe these steps in detail.

\subsubsection{Equivalence Analyzer.}
\label{Se:equiv-analyzer}
The equivalence analyzer is an LLM-based agent equipped with \textit{cat} and \textit{code\_search} tools used to answer the program equivalence queries.
The agent is provided with partial code snippets and an equivalence query
(along with a system prompt, see~\sectref{system-prompts-equiv}).
The agent uses the available tools to obtain the relevant code snippets needed to answer the program equivalence query;
and outputs the relevant code snippets and an explanation of whether the code semantics are preserved or not.
By retrieving the relevant code snippets, the agent does ``implicit slicing'' of the program to the portions is required to answer
the user's equivalence query. We note that this is a particular advantage of an LLM-based approach to answering such queries.

\subsubsection{Formalizer.}
\label{Se:equiv-formalizer}
Given the relevant code snippets from the equivalence analyzer, the \emph{formalizer} produces a code abstraction similar to 
a Program-Dependence-Graph (PDG)~\cite{ferrante1987program}, which is a directed graph representing a program's control and data dependencies with statements as nodes and dependencies as edges.
The abstractions of the code are captured using logical predicates. More details on the abstractions are discussed later in this section.

In this work, we use an LLM as the underlying formalizer. The formalization step
is a call to an LLM with the relevant code snippets (obtained from the analyzer~\sectref{equiv-analyzer}) and a system prompt (see~\sectref{system-prompts-equiv})
with a vocabulary of predicate signatures\footnote{We distinguish between \textit{predicate signatures}, which specify the relation and its arity (e.g., $\tt isConstantValue(x: str)$), and the \textit{predicate instances (facts)}, which apply these signatures to concrete arguments (e.g., $\tt isConstantValue(``0")$). When no ambiguity arises, we use the term ``predicates'' informally to refer to either.} that the model should use to capture the code
semantics and the correspondence between the programs. The model is instructed to explain equivalence using the code snippets
and the logical predicate signatures available.
The model's response contains three parts:

\begin{itemize}
    \item Formal logical predicates (concrete instances) capturing the code semantics
    \item  Required correspondence information between the two programs
    \item  An explanation -- mixing natural language and predicates -- of the equivalence outcome.
\end{itemize}

The first two items (logical predicates and correspondence) constitute the \AgentClaims~in the context of
program equivalence, which are next validated using a \VerificationCondition (see ~\sectref{equiv-verification-condition}). The third item (explanation) is not used for formal verification, and instead is an artifact used to ground the model's output.~\sectref{equiv-explain-artifact} shows an instance of this artifact for a simple example.


\paragraph{\textit{Implementation Details.}}

\begin{table}[t!]
    \centering
    \resizebox{0.99\linewidth}{!}{
    \begin{tabular}{|l|l|}
    \hline
        \textbf{Predicate List} &  \textbf{Semantics}\\
    \hline
    $\tt use(x: str, f: str, l: num)$ & Variable $\tt x$ is used at line $\tt l$ in file $\tt f$.\\
    \hline
    $\tt def(x: str, f: str, l: num)$ & Variable $\tt x$ is defined at line $\tt l$ in file $\tt f$.\\
    \hline
    $\tt flow(x: str, f_1: str, l_1: num, $ & The value of variable $\tt x$ defined at line $\tt l_1$ in file $\tt f_1$ \\
    \quad $\tt y: str, f_2: str, l_2: num)$ & flows to the variable $\tt y$ at line $\tt l_2$ in file $\tt f_2$.\\
    \hline
    $\tt controldep(x: str, f_1: str, l_1: num,$ & Variable $\tt x$ at $\tt l$ in file $\tt f$ is control-dependent\\
    \quad $\tt cond: str, choice: bool, f_2: str, l_2: num)$ & on $\tt cond$ with the $\tt choice$ branch at $\tt l_2$ in $\tt f_2$.\\
    \hline
    \multirow{2}{*}{$\tt defWithExpr(x: str, f: str, l: num)$} & Variable $\tt x$ defined at $\tt l$ in $\tt f$ is a result of a complex expression,\\
    & i.e. it is not a constant or a copy assignment.\\
    \hline
    \multirow{2}{*}{$\tt condWithExpr(f: str, l: num)$} & Indicates a complex conditional at line $\tt l$ in file $\tt f$,\\
    & i.e., not a single variable or a negation of a variable.\\
    \hline
    $\tt unaryFun(operator: str, operand: str,$ & Indicates a unary function with one $\tt operator$\\
    \quad $\tt f: str, l: num)$& and one $\tt operand$ at $\tt l$ in $\tt f$. \\
    \hline
    $\tt binaryFun(op: str, opd_1: str,$ & Indicates a binary function with an operator $\tt op$ and \\
    \quad $\tt opd_2: str, f: str, l: num)$& two operands $\tt opd_1, opd_2$ at $\tt l$ in $\tt f$. \\
    \hline 
    $\tt entry(fun: str, f: str, l: num)$ & Function $\tt fun$ starts at line $\tt l$ in file $\tt f$.\\
    \hline
    $\tt exit(f: str, l: num)$ & The main function returns at line $\tt l$ in file $\tt f$.\\
    \hline
    $\tt isConstantValue(x: str)$ & Variable $\tt x$ is a constant value.\\
    \hline
    $\tt watchVar(x: str, f: str, l: num)$ & Variable $\tt x$ is a variable on which equivalence is checked.\\
    \hline
    $\tt varMap(x: str, f_1: str, l_1: num, $ & Creates a mapping between variables $\tt x$ in Code1 and $\tt y$ in Code2,\\
    \quad $\tt y: str, f_2: str, l_2: num)$ & to capture textually-different but equivalent variables.\\
    \hline
    \multirow{2}{*}{$\tt entryMap(f_1: str, l_1: num, f_2: str, l_2: num) $} & Creates a mapping between the entry points\\
    &  of the main function in Code1 and Code2.\\
    \hline
    \multirow{2}{*}{$\tt exitMap(f_1: str, l_1: num, f_2: str, l_2: num) $} & Creates a mapping between the exit points\\
    &  of the main function in Code1 and Code2.\\
    \hline
    \end{tabular}
    }
    \caption{Vocabulary of predicate signatures to capture the code semantics for program equivalence queries.}
    \label{Ta:predicate-list}
\end{table}

\begin{figure}[t!]
    \centering
\begin{subfigure}[t]{0.48\linewidth}
\begin{minipage}[t]{\linewidth} 
    \begin{lstlisting}
// main.cpp
...
1. int a = 0;
2. int b = 2;
3. bool c = a == b;
4. int d = 2;
5. if (c) {
6.   d = foo(a);
7. }
...
    \end{lstlisting}
\end{minipage}%
\caption{Sample code snippet 1}
\label{Fi:program-predicates-example-code-1}
\end{subfigure}
\hfill 
\begin{subfigure}[t]{0.48\linewidth}
\begin{minipage}[t]{\linewidth} 
    \begin{lstlisting}
// main.cpp
...
1. int a = 0;
2. int b = 2;
3. bool c = a == b;
4. int d = 2;
5. if (c) {
6.   d = bar(a);
7. }
...
    \end{lstlisting}
\caption{Sample code snippet 2}
\label{Fi:program-predicates-example-code-2}
\end{minipage}

\end{subfigure}
\caption{Two code snippets differing on Line 6; $\tt d$ is assigned as a result of $\tt foo$ in Code 1 and $\tt bar$ in Code 2.}
\label{Fi:program-predicates-example}
\end{figure}

\begin{figure}[t!]
    \centering
\begin{subfigure}[t]{0.48\linewidth}
\begin{minipage}[t]{\linewidth} 
    \begin{lstlisting}[language={}]
def("0", 0).
def("2", 0).
def("a", 1).
use("0", 1).
flow ("0", 0, "0", 1).
flow("0", 1, "a", 1).
def("b", 2).
use("2", 2).
flow ("2", 0, "2", 2).
flow("2", 2, "b", 2).
def("c", 3).
use("a", 3).
use("b", 3).
flow("a", 1, "a", 3).
flow("b", 2, "b", 3).
defWithExpr("c", 3).
binaryFun("==", "a", "b", 3).
def("d", 4).
use("2", 4).
flow("2", 0, "2", 4).
flow("2", 4, "d", 4).
use("c", 5).
flow("c", 3, "c", 5).
def("d", 6).
use("a", 6).
defWithExpr("d", 6).
controldep("d", 6, "c", true, 5).
unaryFun("foo", "a", 6).
exit(8).
use("a", 8).
flow("a", 1, "a", 8).
use("b", 8).
flow("b", 2, "b", 8).
use("c", 8).
flow("c", 3, "c", 8).
use("d", 8).
flow("d", 4, "d", 8).
flow("d", 6, "d", 8).
\end{lstlisting}
\end{minipage}%
\caption{Predicates for~\figref{program-predicates-example-code-1}}
\label{predicates-example-code-1}
\end{subfigure}
\hfill 
\begin{subfigure}[t]{0.48\linewidth}
\begin{minipage}[t]{\linewidth} 
    \begin{lstlisting}[language={}]
def("0", 0).
def("2", 0).
def("a", 1).
use("0", 1).
flow ("0", 0, "0", 1).
flow("0", 1, "a", 1).
def("b", 2).
use("2", 2).
flow ("2", 0, "2", 2).
flow("2", 2, "b", 2).
def("c", 3).
use("a", 3).
use("b", 3).
flow("a", 1, "a", 3).
flow("b", 2, "b", 3).
defWithExpr("c", 3).
binaryFun("==", "a", "b", 3).
def("d", 4).
use("2", 4).
flow("2", 0, "2", 4).
flow("2", 4, "d", 4).
use("c", 5).
flow("c", 3, "c", 5).
def("d", 6).
use("a", 6).
defWithExpr("d", 6).
controldep("d", 6, "c", true, 5).
unaryFun("bar", "a", 6).
exit(8).
use("a", 8).
flow("a", 1, "a", 8).
use("b", 8).
flow("b", 2, "b", 8).
use("c", 8).
flow("c", 3, "c", 8).
use("d", 8).
flow("d", 4, "d", 8).
flow("d", 6, "d", 8).
    \end{lstlisting}
\end{minipage}
\caption{Predicates for~\figref{program-predicates-example-code-2}}
\label{Fi:predicates-example-code-2}
\end{subfigure}
\hfill
\begin{subfigure}[t]{\linewidth}
\begin{minipage}[t]{\linewidth} 
    \begin{lstlisting}[language={}]
entryMap("main", 0, "main", 0).
exitMap(8, 8).
    \end{lstlisting}
\end{minipage}
\caption{Predicates capturing a correspondence between programs in~\figref{program-predicates-example}}
\end{subfigure}
\caption{Figure shows the predicates for programs in~\figref{program-predicates-example} according to the predicate list in~\tableref{predicate-list}. Note that, for simplicity, the file name $\tt main.cpp$ is omitted for every predicate in this list.}
\label{Fi:predicates-example}
\end{figure}

~\tableref{predicate-list} shows the vocabulary of predicates used to capture 
the semantics of the code under consideration. The predicates $\tt varMap$, $\tt entryMap$, and $\tt exitMap$ capture explicit correspondences between the two programs. Specifically, $\tt varMap$ maps variables that differ textually but are semantically equivalent; $\tt entryMap$ aligns the entry points (line numbers) of the two programs; and $\tt exitMap$ aligns their exit points. These mappings establish the correspondence between the programs, while the remaining predicates encode their individual code semantics.

In this work, we have normalized functions and expressions
to unary and binary functions before passing them to the formalizer. An expression of the form $\tt x = a + b + c;$ is normalized to $\tt x\_tmp = a + b; x = x\_tmp + c;$. Though we used the normalization technique in this work, one can skip the need for normalization by enriching the formalization step to capture the code semantics more thoroughly and precisely.

\paragraph{Example}
\figref{program-predicates-example} shows two sample, normalized programs that are very similar except for a local and minor change
on Line 6. In Code 1, $\tt d$ is redefined on Line 6 with a call to $\tt foo(a)$ and is redefined with a call to $\tt bar(a)$ in Code 2.~\figref{predicates-example} shows the predicates for the programs capturing the semantics using the predicate list in~\tableref{predicate-list}. We can observe that the expression $\tt a == b$ is captured as a binary function with $\tt == $
as the function symbol or the operator $\tt binaryFun(``==", ``a", ``b", 3)$ and the function call on Line 6 is represented
as a unary function $\tt unaryFun(``foo", ``a", 6)$ in Code 1 and $\tt unaryFun(``bar", ``a", 6)$ in Code 2. Note that the file name $\tt ``main.cpp"$ is removed for simplicity in~\figref{predicates-example}.

\subsection{Design space of predicates for equivalence}
The predicates needed to formulate meaningful agent claims are -- like other program analyses -- determined by the task that the agent is asked to perform. For example, determining variable uninitialization (as done in 
~\sectref{uninit-vars}) required predicates that capture def-use and flow of values along a single stack trace. In contrast, program equivalence requires predicates that can capture every program flow, i.e., every trace or path in the program. We now discuss the design space for these predicates and the challenges that must be addressed to produce a minimal predicate set for the equivalence task.

\paragraph{Modeling Dataflow Dependencies.}
Similar to the modeling of dataflow in the case of uninitialized variables,
we start designing the predicate set for answering program equivalence
by capturing dataflow dependencies using ``def-use'' ($\tt def$, $\tt use$) and ``data flow'' ($\tt flow$) analysis from the program analysis literature that capture the flow of values from one program point to another program point. Note that we model constants as variables defined at the start of the program, typically on Line 0, and these constants flow to different ``uses'' throughout the program.

\paragraph{Modeling Control Dependencies.}
The control dependency information between two program statements is also important to argue about equivalence and is captured using the $\tt controldep$ predicate.

\paragraph{Modeling Expressions and Library Functions.}
Data-flow analysis tracks how values move across program points, but it cannot fully capture \textit{how values are computed}. Expressions with identical data-flow and control-dependencies may differ in operators, operand order, or function calls, leading to different semantics.

To address this, in an initial design, we treated expressions (that are not single variables) and library calls as \textit{uninterpreted, non-deterministic functions}, introducing new predicates ($\tt defWithExpr$ and $\tt condWithExpr$) to denote unique values. By default, program abstractions with such constructs are treated as non-equivalent

To reduce conservativeness, we instead ultimately chose to model expressions as $\tt unaryFun$ or $\tt binaryFun$, where the operator is the function symbol and operands are its arguments. This approach is consistent with many programming languages (e.g., OCaml, Scala)~\cite{ocaml:operators}, where operators are already viewed as functions. Two expressions are then equivalent if they share the same operator and their arguments are equivalent.

\paragraph{Modeling Textual Differences.}
The equivalence analysis can be improved by accounting for textual differences that are semantically equivalent between two programs.
One can use the standard way of variable-renaming or static single assignment~\cite{cytron1991efficiently}.
We address this issue by using a predicate $\tt varMap$ to capture the variable mapping between two variables that are only textually different but semantically equivalent, such as $\tt varMap(x, p_1, y, p_2)$. This information is captured for every variable pair that is only textually different at the point of the variable definition. 

\paragraph{Modeling Variables of Interest.}
As discussed in~\sectref{equiv-verification-condition},
we use predicate $\tt watchVar$ to capture the variables in each program on which program equivalence is assessed. This predicate can be used to list either only certain variables or all variables, depending on the definition of equivalence.

The final step of the pipeline involves formal verification using \Souffle.
The predicates obtained from
the formalizer, i.e., predicates for both programs and a correspondence between the programs, along with the verification condition (as discussed in~\sectref{equiv-verification-condition}), are fed into \Souffle.
The \Souffle engine then checks the predicates against the verification condition and output whether the programs are equivalent or not.

\subsection{Iterative Formalization for Missing Predicate Synthesis}
\label{Se:iterative-missing-predicates}
As discussed earlier, given two programs, an LLM-based agent is used as a formalizer to generate logical predicates (instances like shown in~\figref{predicates-example}) that capture code semantics and provide an explanation using the generated predicates. However, as the code snippets become more complex and the predicate set needed to capture the semantics increases, the agent sometimes fails to include all the required predicates in its response. 

In our observation, when the LLM fails
to use some predicates, these tend to be predicates that are an artifact of our verification condition but are not fundamentally about equivalence. 
Some examples of missing predicates include (a) the agent not generating calls to the $\tt watchVar$ predicate, or (b) the agent skipping the generation of the $\tt use$ and $\tt flow$ predicates for every $\tt watchVar$ predicate (this is primarily required for easy verification using a formal verification tool), etc. To overcome this issue, we iteratively call the formalizer with the relevant code snippets and the predicates generated so far. The formalizer agent is asked if the current predicate instance set (generated so far) sufficiently captures the code semantics to answer the equivalence query, and if not, to generate the predicate instances that ``it thinks'' are missing~\footnote{Note that we are only asking the LLM to generate predicate instances and not new predicate signatures.}.

The predicates from every call to the formalizer are then consolidated by concatenating the predicates, thereby filling the missing predicates.

\paragraph{Stopping Criteria} The formalizer is run for a maximum of five times or when the LLM concludes that the predicates are sufficient and stops generating new predicate instances, whichever is sooner.

\paragraph{Verification Step checks if the generated predicate instances are sufficient.} The verification step checks if the generated predicate instances set is sufficient and outputs ``inconclusive'' if the generated predicates are not sufficient.

\subsection{Evaluation Dataset}
\label{Se:eval-dataset-equiv}
To evaluate our approach, we used a dataset of 20 problems, manually picked, primarily from two different domains:

\begin{figure}[t!]
    \centering
\begin{subfigure}[t]{0.46\textwidth}
    \begin{minipage}[t]{\linewidth} 
    \begin{lstlisting}
// check.h
29|typedef struct {
30| int* arr;
31| int size;
32|} Cache;

// check.cpp
25|int CacheInit(Cache* const cache, int size) {
26| const int cache_size = 1 << size;
27| cache->arr = (int*)malloc(
28| (int)cache_size, sizeof(*cache->arr));
29| if (cache->arr == NULL) {
30|     return 0;
31| }
32| cache->size = size;
33| return 1;
34|}
    \end{lstlisting}
\end{minipage}%
\caption{Code without -fbounds-safety annotation}
\label{Fi:fbounds-example-without-annotation}
\end{subfigure}
\hfill 
\begin{subfigure}[t]{0.51\textwidth}
\begin{minipage}[t]{\linewidth} 
    \begin{lstlisting}
// check.h
29|typedef struct {
30| int* __counted_by_or_null(1 << size) arr;
31| int hash_bits;
32|} Cache;

// check.cpp
25|int CacheInit(Cache* const cache, int size) {
26| const int cache_size = 1 << size;
27| cache->size = cache_size;
28| cache->arr = (int*)malloc(
29| (int)cache_size, sizeof(*cache->arr));
30| if (cache->arr == NULL) {
31|     return 0;
32| }
33| return 1;
34|}
    \end{lstlisting}
\end{minipage}
\caption{Code with -fbounds-safety annotation}
\label{Fi:fbounds-example-with-annotation}
\end{subfigure}
\caption{Two code snippets (a) without -fbounds-safety annotation, (b) with -fbounds-safety annotation are shown. Note that,
compilation with -fbounds-safety flag requires updates to the array and associated counter to be side-by-side and hence the order
of assignments is changed in (b).}
\label{Fi:fbounds-example}
\end{figure}

\begin{itemize}
    \item \textbf{CRQBench (12):} We collected 12 examples from the CRQBench~\cite{dinella2024crqbench} dataset with problems checking for program equivalence.
    CRQBench is a benchmark of 100+ code reasoning questions obtained from contextualized code-review comments, of which we 
    chose 12 sample problems that are of type -- "Is there a difference by changing $\dots$ to $\dots$ in the program. Explain your reasoning." 
    \item \textbf{Fbounds-Safety (7):}  -fbounds-safety~\cite{clangBoundsSafety} is a C extension designed to prevent out-of-bounds (OOB) memory accesses by enforcing bounds safety.
    Along with adding the -fbounds-safety annotation to C code, the compiler, with the -fbounds-safety flag enabled, requires the annotated code to hold certain properties; for example, updates to pointers and associated counters must be made side by side.  As a result, to make the C code memory-safe with the flag, the C code needs to be
    modified slightly, while ensuring that the code semantics are preserved. We selected seven sample programs, with and without annotations and code changes, to verify program equivalence.
    ~\figref{fbounds-example} shows an example with and without transformation with -fbounds-safety annotation.
\end{itemize}

One additional example~\figref{msan-equiv-example} is our own creation taken from an everyday task. Out of the 20 examples, 15 examples are chosen such that the current analysis is sufficient to
completely and correctly capture the code semantics. For the additional five examples, the current
analysis is either insufficient or the agent doesn’t extract enough relevant code to answer the equivalence query correctly.
These five examples are all taken from CRQBench.

\subsection{Results}

The equivalence analyzer is run on these 20 examples.
The equivalence formalizer is then run on the code snippets generated by the
analyzer. The formalizer is run three times on all 20 examples to understand how the formalizer generates predicates across different runs. 
The experiments are run using Gemini 2.5 Pro~\cite{comanici2025gemini} with single sampling, temperature(t) = 0.2 and top\_p = 0.95 configuration.


As discussed in~\sectref{iterative-missing-predicates}, we run the formalizer iteratively till (hopefully) all the predicates required to capture code semantics are generated. We observed that the agent always generates all the required predicate instances across three runs. From hereon, we will only discuss the output of a single run because we observed that every run leads to the same output.

\begin{figure}
    \centering
    \includegraphics[width=0.7\linewidth]{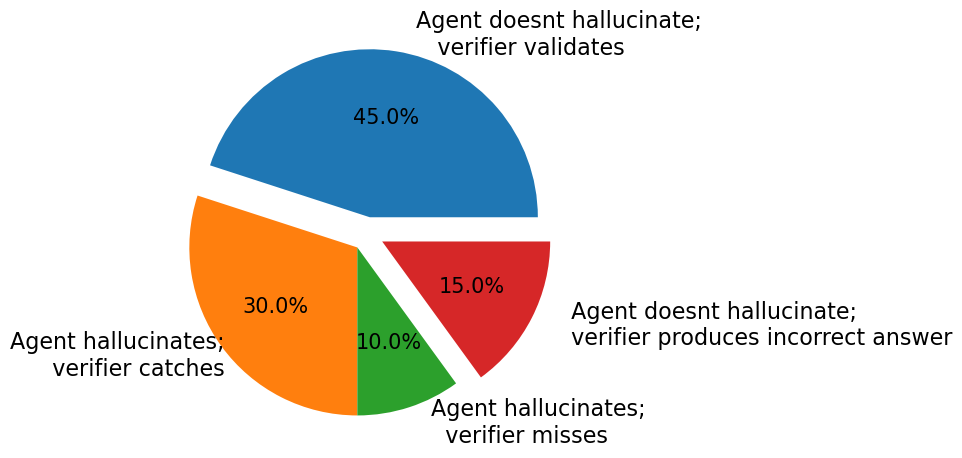}
    \caption{Figure shows the fraction of problems from each category based on the agent's response and the verifier's output. Notice that the agent hallucinates on 40\% of the problems (8/20), with 75\% of them being caught by the verifier.
    }
    \label{Fi:results-pie}
\end{figure}

\figref{results-pie} shows the fraction of problems from each category based on the agent's response and the verifier's output for the 20 examples. Each category is as follows (A = Agent Claims, V = Verifier):

\begin{itemize}
    \item \textbf{A=no hallucination; V=validates}: The agent doesn't hallucinate; the verifier validates the agent's response.
    \item \textbf{A=hallucinates; V=catches}: The agent hallucinates; but the verifier catches the hallucinations.
    \item \textbf{A=hallucinates; V=misses}: The agent hallucinates; but the verifier misses catching the hallucinations.
    \item \textbf{A=no hallucination; V=incorrect output}: The agent doesn't hallucinate, but the verifier doesn't validate the agent's response; the verifier outputs a wrong answer.
\end{itemize}

The percentage of examples where the agent does not hallucinate and the verifier validates the result (blue sector: “A=no hallucination; V=validates”) is $9/20 = 45\%$. The agent hallucinates in $8/20 = 40\%$ of the examples (orange: “A=hallucinates; V=catches” and green: “A=hallucinates; V=misses”), of which the verifier correctly flags $6/8 = 75\%$ (orange). Thus, overall, the verifier catches $6/20 = 30\%$ of the agent’s errors. For the remaining two cases, the errors could be mitigated with improved prompting or stronger models. Finally, in $3/20 = 15\%$ of the examples (red: “A=no hallucination; V=incorrect output”), the agent doesn't hallucinate but verifier, though sound, itself produces an incorrect result due to the incompleteness of the current analysis. These errors can be reduced by enriching the vocabulary of the predicates, leading to more precise analysis (discussed later in detail).

We now explore each of these outcomes in more detail.

\begin{table}[t!]
    \centering
    \begin{tabular}{|c|c|c|c|}
        \cline{3-4}
        \multicolumn{2}{c}{} & \multicolumn{2}{|c|}{Agent's Response (Agent Claims)}  \\
        \cline{3-4}
         \multicolumn{2}{c|}{} & YES & NO\\
         \hline
         \multirow{2}{*}{Ground Truth} & YES & True Positive (3) & False Negative (0) \\
         \cline{2-4}
         & NO & False Positive (6) & True Negative (6)\\
         \hline
    \end{tabular}
    \caption{Agent's Response vs Ground Truth for the 15 examples for which the predicate set is sufficient to capture the semantics. \textit{The number in the bracket for every category indicates the number of examples that fall into that category.}}
    \label{Ta:results-1}
\end{table}

We begin with the 15 examples where our analysis successfully captures the code semantics, and then turn to the remaining 5 examples (see~\sectref{eval-dataset-equiv} for details).

\paragraph{\textbf{A=no hallucination; V=validates}}
For the question of ``Are programs equivalent?'' for the 15 examples,
~\tableref{results-1} shows the agent's response with respect to the ground truth. We can observe that the agent’s response is consistent with the ground truth in a total of 12 cases -- 3 examples where the programs are equivalent and seven examples where they are not.
In all 12 examples, the verification step output is consistent with the agent’s claims and the ground truth, and this provides the user confidence that the agent has substantiated its answer in a logically consistent way. The number in the bracket for every category indicates the number of examples that fall
into that category.

\paragraph{\textbf{A=hallucinates; V=catches}}
In the case of the six examples where they are ``False Positives'', the agent claims that the
programs are equivalent, whereas in reality, the programs were not equivalent. This is
particularly problematic as the agent hallucinates and falsely outputs that the programs are equivalent.
In all six examples, the verification step was able to catch these hallucinations and
output that the programs are not equivalent. This enables users to validate the agent’s output instead of relying on it without scrutiny.

The examples where the agent hallucinates fall into three categories:
\begin{enumerate}
    \item The agent makes unwarranted assumptions about library functions.
    \item The agent makes assumptions (that may or may not be true) beyond the provided code snippets.
    \item The agent hallucinates about the code semantics.
\end{enumerate}

\subsubsection{The agent makes unwarranted assumptions about library functions.}
~\figref{program-equivalence-example} shows an example where the agent makes unwarranted assumptions about library functions. The agent incorrectly assumes that $\tt has\_value$ and $\tt !IsEmpty$ are equivalent and as a result, variables $has\_value$ and $\tt is\_empty$ are equivalent. Thereby, the agent concludes that the programs are equivalent. However, in reality (on manual inspection), the functions are not semantically equivalent, and as a result, the programs are not equivalent. 

Because the verification step captures the functions as uninterpreted functions and
conservatively declares them are not-equivalent due to differing function names, the verification
step outputs that the programs are not equivalent and thereby catches such hallucinations
made by the agent.

\subsubsection{The agent makes assumptions (that may or may not be true) beyond the provided
code snippets.}
~\figref{fbounds-example} shows an example where the agent makes assumptions about program behavior beyond the provided code snippets.
In~\figref{fbounds-example-without-annotation}, the variable $\tt cache$->$\tt size$ on Line 32 is assigned \textbf{after} the condition on Line 29, and in~\figref{fbounds-example-with-annotation}, the same variable is defined on Line 27 \textbf{before} the condition on Line 30.
That is, the variable $\tt cache$->$\tt size$ assignment in the first program is control-dependent on the condition $\tt cache$->$\tt arr == NULL$ evaluating to $\tt false$, but the same variable does not have the same control dependency in the second program, i.e., when the programs return on Line 30 and Line 31 respectively, the assignment of the variable $\tt cache$->$\tt size$ in both programs is different. As a result, the two programs are not equivalent. 

However, the agent assumes that the caller function would appropriately handle the variable assignment $\tt cache$->$\tt size$ when the two programs return on Lines 30 and 31, respectively, ensuring that the program semantics remain the same. Based on this assumption, the agent incorrectly concludes that the programs are equivalent.

The verification step determines that the code snippets are not equivalent by identifying the difference in the $\tt controldep$ predicates generated by the agent.

\subsubsection{The agent hallucinates about the code semantics.}

\begin{figure}[t]
    \centering
\begin{subfigure}[t]{0.48\textwidth}
    \begin{minipage}[t]{\linewidth} 
    \begin{lstlisting}
// main.cpp
5| absl::StatusOr<unsigned> find(unsigned start, char *str, char goal) {
6| unsigned len = strlen(str);
7| for (unsigned i = start; str[i] != 0 && i < len; i++) {
8| char *tmp = str;
9| if (tmp[i] == goal) return i;
10| }
11| return absl::NotFoundError("goal not found");
12| }
    \end{lstlisting}
\end{minipage}%
\caption{Code with memory access issue.}
\label{Fi:msan-bug-example}
\end{subfigure}
\hfill 
\begin{subfigure}[t]{0.48\textwidth}
\begin{minipage}[t]{\linewidth} 
    \begin{lstlisting}
// main.cpp
5| absl::StatusOr<unsigned> find(unsigned start, char *str, char goal) {
6| unsigned len = strlen(str);
7| for (unsigned i = start; i < len && str[i] != 0; i++) {
8| char *tmp = str;
9| if (tmp[i] == goal) return i;
10| }
11| return absl::NotFoundError("goal not found");
12| }
    \end{lstlisting}
\end{minipage}
\caption{Code without memory access issue.}
\label{Fi:msan-nobug-example}
\end{subfigure}
\hfill 
\begin{subfigure}[t]{0.98\textwidth}
\begin{minipage}[t]{\linewidth} 
    \begin{lstlisting}[language={}]
The two code snippets are semantically equivalent.
...
5. **Equivalence**:
* In the **Original Code**, the loop continues if `str[i] != 0` AND `i <
len`. When `i` becomes equal to `len`, `str[i]` is `\0`, so `str[i] != 0`
becomes false, and the loop terminates.
* In the **Modified Code**, the loop continues if `i < len` AND `str[i]
!= 0`. When `i` becomes equal to `len`, the first condition `i < len` becomes
false. Due to short-circuiting, the second condition (`str[i] != 0`) is not
evaluated, and the loop terminates.
In both scenarios, the loop iterates over the exact same range of indices (from
`start` to `len - 1`). The termination condition is met at the same point (`i
== len`).
    \end{lstlisting}
\end{minipage}
\caption{Excerpt from agent's explanation for the equivalence query of the programs.}
\label{Fi:msan-equiv-explanation}
\end{subfigure}

\caption{Two code snippets (a) with a memory access issue, (b) without any memory access issue are shown. The swap of conditions $\tt str[i] != 0$ and $\tt i < len$ in (a) causes the memory issue. (c) shows an excerpt from the agent's explanation on why the programs are equivalent.}
\label{Fi:msan-equiv-example}
\end{figure}

~\figref{msan-equiv-example} shows two programs in which the condition in $\tt for$ loop on Line 7 are swapped. As a result, the program in~\figref{msan-bug-example} has a memory access bug, and the program in~\figref{msan-nobug-example} has no memory issues. When the agent is asked if the two programs are equivalent, it hallucinates that the code snippets are equivalent, as shown in~\figref{msan-equiv-explanation}.
The agent fails to consider all scenarios and falsely concludes that the programs are equivalent.

The verification step outputs that the code snippets are \textbf{not equivalent} by using the $\tt controldep$ predicates that capture the difference in control-dependency information, thereby catching the agent's hallucinations.

\paragraph{Additional five examples where the current analysis is insufficient.}
As discussed earlier in the section, five examples are such that either the agent doesn’t fully retrieve the
relevant code snippets to correctly answer equivalence, or the predicates are not precise enough
to capture the information.

\begin{table}[t!]
    \centering
    \begin{tabular}{|c|c|c|c|}
        \cline{3-4}
        \multicolumn{2}{c}{} & \multicolumn{2}{|c|}{Agent's Response (Agent Claims)}  \\
        \cline{3-4}
         \multicolumn{2}{c|}{} & YES & NO\\
         \hline
         \multirow{2}{*}{Ground Truth} & YES & True Positive (3) & False Negative (2) \\
         \cline{2-4}
         & NO & False Positive (0) & True Negative (0)\\
         \hline
    \end{tabular}
    \caption{Agent's Response vs Ground Truth for five examples where the current analysis is not sufficient to answer the equivalence query correctly. \textit{The number in the bracket for every category indicates the number of examples that fall into that category.}}
    \label{Ta:results-2}
\end{table}

\begin{table}[t!]
    \centering
    \begin{tabular}{|c|c|c|c|}
        \cline{3-4}
        \multicolumn{2}{c}{} & \multicolumn{2}{|c|}{Verification Output}  \\
        \cline{3-4}
         \multicolumn{2}{c|}{} & YES & NO\\
         \hline
         \multirow{2}{*}{Agent Claims} & YES & True Positive (0) & False Negative (3) \\
         \cline{2-4}
         & NO & False Positive (0) & True Negative (2)\\
         \hline
    \end{tabular}
    \caption{Agent Claims vs Verification Output for five examples where the current analysis is not sufficient to answer the equivalence query correctly. \textit{The number in the bracket for every category indicates the number of examples that fall into that category.}}
    \label{Ta:results-3}
\end{table}

~\tablerefs{results-2}{results-3} show two tables that highlight the agent claims vs ground truth and verification output vs ground truth for the five examples.

\begin{figure}[t]
    \centering
\begin{subfigure}[t]{0.48\textwidth}
    \begin{minipage}[t]{\linewidth} 
    \begin{lstlisting}
// main.cpp
147: // 3)
148: switch (checker) {
149: case A:
150:    counter->Increment("A");
151:    return "A1";
152: case B:
153:    counter->Increment("B");
154:    return "B1";
155: default:
156:    break;
157: }
158:
159: // 4)
160: counter->Increment("C");
161: return "C1";
    \end{lstlisting}
\end{minipage}%
\caption{Code snippet with $\tt switch$ statement.}
\label{Fi:switch-example}
\end{subfigure}
\hfill 
\begin{subfigure}[t]{0.48\textwidth}
\begin{minipage}[t]{\linewidth} 
    \begin{lstlisting}
// main.cpp
147: // 3)
148: if (checker == A) {
149:    counter->Increment("A");
150:    return "A1";
151: } else if (checker == B) {
152:    counter->Increment("B");
153:    return "B1";
154: }
155:
156: // 4)
157: counter->Increment("C");
158: return "C1";
    \end{lstlisting}
\end{minipage}
\caption{Code snippet with $\tt if-else$ statement.}
\label{Fi:if-else-example}
\end{subfigure}
\caption{Two semantically equivalent code snippets are shown where (a)  uses a $\tt switch$ statement and (b) uses an $\tt if-else$ statement.}
\label{Fi:switch-if-else-example}
\end{figure}

\begin{figure}[t]
    \centering
\begin{subfigure}[t]{0.48\textwidth}
    \begin{minipage}[t]{\linewidth} 
    \begin{lstlisting}
// main.cpp
161: // Make the Request.
162: response =
163: fetcher.value()->Fetch(request);
164:
165: if (!response.ok()) {
166: absl::PrintF("Error");
167: return EXIT_FAILURE;
168: }
    \end{lstlisting}
\end{minipage}%
\caption{Code snippet with a call to $\tt value$ function.}
\label{Fi:value-example}
\end{subfigure}
\hfill 
\begin{subfigure}[t]{0.48\textwidth}
\begin{minipage}[t]{\linewidth} 
    \begin{lstlisting}
// main.cpp
161: // Make the Request.
162: response =
163: fetcher->Fetch(request);
164:
165: if (!response.ok()) {
166: absl::PrintF("Error");
167: return EXIT_FAILURE;
168: }
    \end{lstlisting}
\end{minipage}
\caption{Code snippet with no call to $\tt value$ function.}
\label{Fi:no-value-example}
\end{subfigure}

\caption{Two code snippets are shown where (a) has a call to a $\tt value$ function and (b) doesn't.}
\label{Fi:value-no-value-example}
\end{figure}

\paragraph{\textbf{A=no hallucination; V=incorrect output}}
Of the five examples, there are three examples where the agent claims are consistent with the ground truth that the programs are equivalent. However, for the same examples, the verification step output says that the programs are not equivalent.
This is because the current predicate set is not precise enough to capture this equivalence
information.

In example~\figref{switch-if-else-example}, a $\tt switch$ statement in~\figref{switch-example} is changed to $\tt if-elseif-else$ statement in~\figref{if-else-example}
and the agent is asked if the programs are equivalent, and to generate code snippets for the
same. Though the agent correctly answers equivalence, the predicates generated are not
precise enough to compute equivalence in \Souffle.
For example, the agent generates a predicate that Line 153 control-depends on Line 152, but
according to the provided predicate list, it generates a ``use'' predicate for variable $\tt checker$ at Line 148
which has ``no relation'' to the Line 152 that is used in computing equivalence. As a result, \Souffle outputs that the programs are not equivalent.

\paragraph{\textbf{A=hallucinates; V=misses}}
The remaining two examples fall into the category where the agent claims that the
programs are not equivalent, but the ground truth states that the programs are
equivalent. In these cases, the verification step also concludes that the programs are not equivalent. This is because the agent doesn’t retrieve all the relevant code snippets
(functions, callers, etc.) to answer the equivalence query.

~\figref{value-no-value-example} shows two programs with the only difference being the call to the method $\tt value$. The agent
doesn’t retrieve the function body for $\tt value$ and, as a result, concludes that the programs are
not equivalent. The verification step also shows that the programs are not equivalent,
conservatively. However, the ground truth check shows that the programs are indeed equivalent.

\subsubsection{Ablations:}
\begin{figure}
    \centering
    \includegraphics[width=0.7\linewidth]{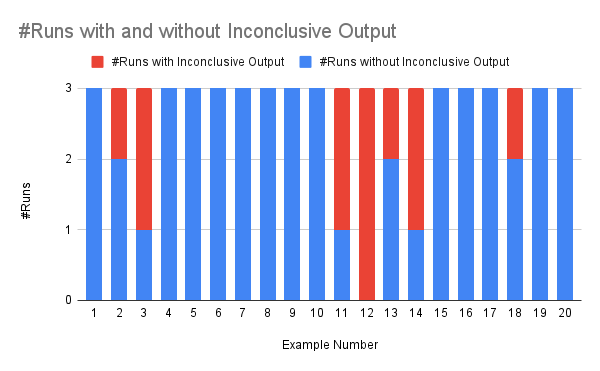}
    \caption{Figure shows the number of times the agent doesn't generate all the required predicates leading to an ``Inconclusive'' output by the verifier.}
    \label{Fi:inconclusive-chart}
\end{figure}
We evaluate the effect of using an iteratively running the formalizer to generate missing predicates. We run the formalizer in two configurations:
\begin{itemize}
    \item Mode 1: Formalizer run once to generate predicates
    \item Mode 2: Formalizer re-run iteratively (a maximum of five times) to generate missing predicate
\end{itemize}
As discussed earlier, on running the agent in Mode 2, the agent always generates all the required
predicates. 
In the case of Mode 1, ~\figref{inconclusive-chart} shows the number of times the agent fails to generate all the required predicates on a single run. We can observe that there are seven examples with at least one run where the agent doesn’t
generate all the predicates necessary, leading to an ``inconclusive'' state. On running the
agent more times (in Mode 2), all the required predicates are then generated, and the verification output across
all runs turns out to be consistent.
This concludes the necessity of running the agent in Mode 2, i.e., running formalizer iteratively to generate missing predicates.

\section{Threats to Validity and Discussion}
\label{Se:future}


It is worth keeping in mind that we inherit the limitations of static analysis via Datalog.  Thus, we cannot circumvent the undecidability of program equivalence, just because we are using LLMs!  To remind the reader, what we are doing in this work is to simply act as a \textit{post facto} check to what an LLM says in its reasoning.  Currently, the system does it well in some cases and not in others, but we believe that the overall "trust" can be increased in this way.


Engineers are increasingly relying on LLMs to explain code to them, and therefore, a high degree of trust in those explanations is crucial. We have presented an approach to reducing the instances of misleading information going to the user. 
However,
it still suffers from false positives and false negatives, and more work is required to minimize these, so engineers can have more trust in LLM's reasoning about code.  We expect improvements needed, for instance, in the design of predicates, in prompts to generate these predicates, and perhaps in an ability to select more powerful verifiers when the reasoning needs exceed the power of Datalog itself.  There are also possibilities of (1) asking the LLM to revise its explanation upon verification failure, and (2) even of fine-tuning the LLM to produce more verifiable explanations in the first place, which should be of great interest to providers of LLMs.

\section{Related Work}
\label{Se:related}

\textbf{Reasoning capabilities of LLMs.} There is a growing body of research focused on improving the reliability of LLMs, particularly in quantitative reasoning domains. A prominent approach involves autoformalization, where an LLM translates informal problem statements into a formal language that can be machine-verified. This is primarily done by prompting an LLM to generate Isabelle code for mathematical problems for automatic verification~\cite{zhou2024don}. Similarly,~\cite{jiang2022draft} uses LLMs to map informal proofs to formal sketches, which then guide automated theorem provers. Other works, such as~\cite{lalwani2024autoformalizing,kesseli2025logic} focus on converting natural language to first-order logic and using SMR solvers for logical fallacy detection. 
Another work, CODEPLAN~\cite{wen2024unlocking}, focuses on encoding the reasoning tasks in the form of \textit{code plans} to capture rich semantics and perform reasoning tasks. 
They demonstrate that using a codeplan improves the reasoning abilities
on a set of (non-code) reasoning tasks.
Our work builds on this paradigm, but instead of focusing on mathematical or general logical reasoning, we apply the principle of formal verification to the domain of code reasoning.
\cite{NEURIPS2024_48bb60a0} is another analogous work on the ``verification'' aspect, but in the context of transpilation or lifting.

\noindent
\textbf{Code Execution Reasoning Methods.} There have also been works on code execution reasoning with the development of frameworks like CodeMind~\cite{liu2024codemind} that evaluate LLMs' capabilities for code reasoning on various tasks such as output prediction and specification adherence. ExeRScope~\cite{liu2025tool} uses static analysis 
to correlate model errors with the code's syntactic properties, such as program complexity or recursion depth.
Other methods, such as NExT~\cite{ni2024next} aim to directly enhance these reasoning capabilities by fine-tuning models on execution traces
paired with natural language rationales.~\cite{yang2025code} shows techniques that demonstrate (1) how code helps with reasoning in LLMs and (2) how reasoning helps to improve code generation.

\noindent
\textbf{Benchmarks.} There also have been other benchmarks that try to evaluate the reasoning capabilities of LLMs. CodeMMLU~\cite{manh2024codemmlu} demonstrates this using a benchmark of multiple-choice questions covering the topics of code comprehension and software development. CRUXEval~\cite{gu2024cruxeval} measures the execution ability of LLMs, by containing a benchmark of 800 functions with input-output pairs with the goal of producing the output given the input and produces an input that matches the output for the function.
\cite{thakur2025clever},~\cite{lohn2024minicodeprops} are benchmarks aimed at generating proofs in LEAN for programs and specifications.
Another benchmark~\cite{wei2025equibench} introduces 2400 program pairs for evaluating LLMs on equivalence checking. We use~\cite{dinella2024crqbench} that consist of 100+ code-reasoning questions derived from contextualized code review comments.

\noindent
\textbf{Program Equivalence.} A widely studied topic in the program analysis and verification literature is program equivalence. Early works on defining and analysis equivalence use symbolic techniques~\cite{horwitz1990identifying,ferrante1987program,horwitz1992use}, bisimulation methods (survey:~\cite{garavel2022equivalence}), translation validation~\cite{kundu2009proving,tate2009equality} and other formal techniques. More recently LLMs have been used for the task of program equivalence as shown in~\cite{allamanis2025disproving}.

\section{Conclusion}
\label{Se:conclusion}

In this work, we introduced a novel method to automatically validate 
a code reasoning agent's answers by formalizing and verifying their reasoning steps. 
We described the overall framework of extracting \AgentClaims~(a formal representation of the agent's response)
and verifying it against a \VerificationCondition~in a formal, deterministic tool.
We showed our approach in the context of uninitialized variables detected by sanitizers
and program equivalence queries.
For the benchmark of 20 uninitialized variables, the formal verification step was able to validate
the agent's reasoning on 13/20 examples.
For a set of 20 program equivalence queries, taken primarily from CRQBench and Fbounds-Safety, we found that our approach successfully caught 6/8 incorrect agent judgments and validated the agent's response in 9/20 cases. For the remaining five cases, the analysis and model prompting can be improved, leading to better results.

\bibliographystyle{plain}
\bibliography{refs}

\clearpage
\appendix
\section{Appendices}
\subsection{System Prompt for Basic Code Reasoning Agent}
\label{Se:basic-system-prompt}
\figref{agent_prompt} shows the system prompt that was used for a basic code reasoning agent. Here are  a few insights about the prompt:
\begin{itemize}
    \item Asking the model to include relevant code snippets in the prompt is necessary for the subsequent trace extraction steps.
    \item It is essential to keep the examples in the system prompt as generic as possible. Otherwise, the model loses generality.
\end{itemize}

\begin{figure}[t!]
\begin{lstlisting}[language={}]
The allocation of `op` happens on line 272 in `LogTest_BadBytes_Test::TestBody()`:
````c++
272|  Log op;
````c++
This `Log` object `op` is used in the `LogTest::BadBytes` test case. The `DoWork` method of this `op` object is called on line 285:

...
To fix this, Log::Init needs to be called before Log::Start in the `LogTest::BadBytes` test.
\end{lstlisting}
    \caption{Figure shows an excerpt from the agent response for a code-reasoning question.}
    \label{Fi:agent-response-excerpt}
\end{figure}
\figref{agent-response-excerpt} shows an excerpt from the agent's extracted response for a code-reasoning question.

\begin{figure}[t!]
\begin{lstlisting}[language={}]
Your task is to answer a question related to code that a software developer is
working on.

Guidelines:
 1. Think step-by-step.
 2. You have not seen this code. You *must* confirm all your assumptions by
    examining the source code using the available tools. You *must* use the
    `code_search` tool or the `cat` tool to find all the relevant source code to
     answer the questions. For `cat`, you do not need the google3 prefix.

Output instructions:
 1. Explain what you are doing at every step in your response.
 2. Include relevant code snippets and refer to them in your explanation.
 3. Be explicit about the program variables, members, and functions in your
    explanations.
 4. You *must* include the filename in your code snippets.
 5. Be explicit about the line numbers in your explanations.
 6. It is very important to include code snippets. Do not forget that.

Example #1:

I have retrieved the code for `foo.cc`. There is a use of uninitialized value at
line 106. Either `i` or `j` is uninitialized.

```c++
  // foo.cc
  106|   some_fun(i, j);
```
I need to look at how i and j are defined and if there are any subsequent
initializations.

Example #2:

I have retrieved the code for bar.cc. From the code, I can see that 'j' is
initialized, but 'i' is not initialized after line 7 of bar.cc:

```c++
  // bar.cc
  6|   int i;
  7|   int j = 10;
```

\end{lstlisting}
\caption{System prompt for the Code Reasoning Agent.}
\label{Fi:agent_prompt}
\end{figure}

\subsection{More details on errors in LLM reasoning}
\label{Se:program-equivalence-example-elaborate-section}
We provide further experiments illustrating systematic errors in LLM reasoning on program equivalence.

\begin{figure}[t!]
    \centering
    \noindent 
\begin{subfigure}[t]{0.475\linewidth}
    \begin{minipage}[t]{\linewidth} 
        \begin{lstlisting}
// main.cpp
...
80: bool has_value =
81:       session_input.has_value();
82: if (has_value) {
83:   status.check = 0;
84:   status.type =
85:     GetType(session_input.value());
86: }
...

2: constexpr bool has_value() const { 
3.  return this->has_data; 
4. }
        \end{lstlisting}
    \end{minipage}%
    \caption{Code Snippet 1 using \texttt{has\_value} method; additionally the function for \texttt{has\_value} is provided.}
    \label{Fi:program-equivalence-example-elaborate-a}
\end{subfigure}
\hfill 
\begin{subfigure}[t]{0.495\linewidth}
    \begin{minipage}[t]{\linewidth}
        \begin{lstlisting}
// main.cpp
...
80: bool is_empty =
81:       IsEmpty(session_input);
82: if (!is_empty) {
83:   status.check = 0;
84:   status.type =
85:     GetType(session_input.value());
86: }
...

4. bool IsEmpty(const Message& message){ 
5.  return message.ByteSizeLong() == 0; 
6. }
        \end{lstlisting}
    \end{minipage}
    \caption{Code Snippet 2 using \texttt{IsEmpty} method; additionally the function for \texttt{IsEmpty} is provided.}
    \label{Fi:program-equivalence-example-elaborate-b}
\end{subfigure}
\vspace{5ex}
\begin{subfigure}[t]{0.99\linewidth}
    \begin{minipage}[t]{0.99\linewidth}
        \begin{lstlisting}[language={}]
The two code snippets are semantically equivalent.
...
3.  **Comparison**:
    *   The condition to enter the `if` block in the **original code** is `session_input.has_value() == true`, which means the session has data.
    *   The condition to enter the `if` block in the **modified code** is `!IsEmpty(session_input)`, which also means the session has data.

Since the condition for executing the code block on lines 83-86 is identical in both versions (i.e., the `session_input` object is not empty), and the code within the block is unchanged, the two snippets are functionally equivalent. The change is a refactoring from a member method (`has_value`) to a free function (`IsEmpty`) with the corresponding logical inversion at the call site.
        \end{lstlisting}
    \end{minipage}
\caption{Incorrect LLM explanation for program equivalence query; the agent makes incorrect assumptions about \texttt{has\_value} and \texttt{IsEmpty} semantics.}
\label{Fi:program-equivalence-example-elaborate-c}
\end{subfigure}
\vspace{-7ex}
\caption{The two code snippets from~\figref{program-equivalence-example} are shown with the definitions for $\tt has\_value()$
and $\tt IsEmpty()$ where the condition at Line 82 is obtained by checking for ${\tt has\_value()}$ 
in Code 1 and ${\tt IsEmpty()}$ in Code 2.}
\label{Fi:program-equivalence-example-elaborate}
\end{figure}

\sectref{errors-library} discusses a scenario where the agent makes unwarranted assumptions about library behavior. Even after the agent is provided with the definitions for 
$\tt has\_value()$ and $\tt IsEmpty()$ as shown in~\figref{program-equivalence-example-elaborate}; the agent continues to
hallucinate and incorrectly reason about the program equivalence.

\subsection{System Prompts for MSAN Uninitialized Variables Query}
\label{Se:system-prompts-msan}
This section shows the prompts used in the uninitialized memory sanitizer tasks.
\begin{figure}[t!]
\begin{lstlisting}[language={}]
Guidelines:
* You will be given an explanation of why a memory error happened.
* You will also be given the relevant code snippets.
* As a first step, associate all the text in the explanation to the given
  code snippets.
* After associating the explanations to the code snippets, extract a trace
  that shows how the memory error manifests itself.
* The explanation may gloss over some details of the control and data
  flow. Use the code snippets to fill in any missing links in the data
  and control flow so that the trace is complete.
* Make sure you extract all information about data flow and control flow
  in the explanation and associate with the code lines in the trace.
* Show how the data flows between the lines in the trace in terms of the
  variables from the code snippets.
* Show the actual lines from the code snippet.

Example:

  //  `x` is being used.
  //  a/b/c/d.cc
  303 |   foo(x)
  ...
  //   x flows to y, x flows to z
  350 |   y, z  = x

  //  `x` flows to `a`
  //  another/file/fork.cc
  923 |  foo(int a) {{
  924 |  ...
  //  'a' flows to 'y'
  925 |     y = a
  ...
  // y is read here and causes memory error
  949 |     read(y)

FILE_CONTEXT:
  {file_context}
EXPLANATION:
  {explanation}
\end{lstlisting}
\caption{System Instruction for extracting a trace for MSAN bugs from agent's explanation.}
\label{Fi:trace-extraction-system-prompt}
\end{figure}

~\figref{trace-extraction-system-prompt} shows the system prompt for extracting a code trace from the agent's output for MSAN tasks.

\begin{figure}[t!]
\begin{lstlisting}[language={}]
You will be given a code trace along with some text associated with the
code lines. Your task is to convert the text lines in conjunction with the
code snippets into datalog.

Guidelines:
* Go line-by-line.
* Make sure you create the correct `flow` and `use` predicates so that
  the dataflow in the trace is explicitly captured.
* Consolidate all the datalog predicates and output them in the following
  format:
    ``datalog
...
    ``

You can use the following predicates:
* uses("x", "source", line): expression `x` is used at location "source:line".
* uninitialized("x", "source", line): expression `x` is uninitialized at location
  "source:line" and may contain unknown or garbage values.
* hasInitializer("f", "m"): Variable "f" has an initializer in "m".
* hasMemberInitializer("f", "m"): Constructor "m" has an initializer for member variable "f".
* allocated("x", "source", line): Memory referred to by "x" is allocated at
  location "source:line".
* declared("x", "source", line): Variable "x" is declared at location
  "source:line".
* flow("x", "src_file_x", line_x, "y", "src_file_y", line_y): The value of `x` at
  defined at "src_file_x:line_x" flows to the "y" at "src_file_y:line_y". A Flow is
  usually a result of definition reaching a use, an argument being passed at a
  call site,  assignments, return values, etc.
* memoryError("x", "kind", "source", line): There is a `kind` memory error
  involving `x` at location "source:line".

TRACE:
{trace}

\end{lstlisting}
\caption{System Instruction for generating formal representation of a trace for MSAN tasks.}
\label{Fi:formal-representation-system-prompt}
\end{figure}

~\figref{formal-representation-system-prompt} shows the system instruction for generating a formal representation of a trace for MSAN tasks. The system instruction contains a vocabulary of formal predicates to use when generating a formal representation of the trace.

\subsection{Supplementary Experiments for MSAN Bugs}
\label{Se:msan-eval-appendix}

\subsubsection{Manual Judgment of Code Reasoning Agent Output}
In this section, we briefly discuss some additional experiments and analysis of using our approach for MSAN bugs.

\begin{figure}[t!]
  \centering
  \includegraphics[width=0.8\textwidth]{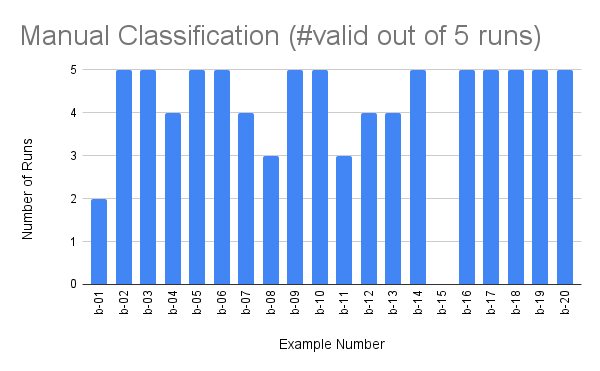}
  \caption{Manual classification of the validity of agent's explanation. }
  \label{Fi:msan_manual_classification}
\end{figure}

As discussed in~\sectref{results-msan}, we ran the code reasoning agent on 20 MSAN examples five times and manually evaluated whether the agent's response is valid or invalid. During the evaluation, a human expert reviewed the agent's trajectories and cross checked it against the underlying code and error message.~\figref{msan_manual_classification} shows the output of the manual evaluation. 

Out of the 100 runs of the agent, 88 runs had a valid explanation based on human judgment, as shown in~\figref{msan_manual_classification}.
The code reasoning agent was able to generate a valid explanation for the vast majority of the MSAN bugs, which is not entirely surprising, because, as explained earlier, this is a simplified version of the code reasoning problem. Nevertheless, MSAN bugs serve as a useful dataset to check the feasibility of the approach. Moreover, even in the simplified problem, the agent makes mistakes for some bugs, and therefore, having an automated check of the agent’s response is still valuable.  


\subsubsection{Automatic validation using Gemini 2.5 Pro}
\begin{figure}[t!]
  \centering
  \includegraphics[width=0.8\textwidth]{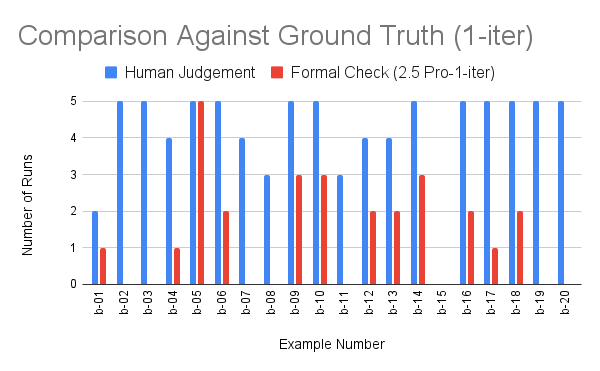}
  \caption{Results of automated validation using Gemini 2.5 Pro with 1 iteration of the formalization step.}
  \label{Fi:msan_pro2.5_1_iter_comparison}
\end{figure}

As an additional experiment, we also performed the formalization step using Gemini 2.5 Pro, and the results are much better even for a single step. In particular, the Gemini 2.5 Pro model was able to prove the validity of 27 out of 88 valid explanations even with a single iteration, as shown in \figref{msan_pro2.5_1_iter_comparison}.

\subsection{System Prompts for Program Equivalence Query}
\label{Se:system-prompts-equiv}
This section shows the system prompts used in the program equivalence tasks.

\lstref{system-inst-1} shows the system instruction given to the Equivalence Analyzer
to extract relevant code snippets and an explanation if the programs are equivalent or not.

The system instruction for the Formalizer to generate the predicates for the code snippets and an explanation of equivalence using the generated predicates is shown in~\lstref{system-inst-2}.

\FloatBarrier
\begin{lstlisting}[caption={System Instruction for Equivalence Analyzer to output the relevant code snippets and an explanation if the programs are equivalent or not.}, label={lst:system-inst-1},language={}]
Additional Instructions:
  * Think step-by-step.
  * You have not seen this code. You *must* confirm all your assumptions by
      examining the source code using the available tools. You *must* use the
      `code_search` tool or the `cat` tool to find all the relevant source code to
      answer the questions. For `cat`, you do not need the google3 prefix.
  * If you want to see the full contents of a file use `cat` rather than code
    search.
  * Provide the relevant code snippets for the question.
  * To answer the user query and provide original and modified code snippets,
    you will have to retrieve code using tools.
  * Explain what you are doing at every step in your response.
  * Include relevant code snippets and refer to them in your explanation.
  * Be explicit about the program variables, members, and functions in your
      explanations.
  * You *must* include the filename in your code snippets.
  * Be explicit about the line numbers in your explanations.
  * It is very important to include code snippets. Do not forget that.
  * Get as much code as possible that will help you to answer the query.
  * For the code change requested in the query, create a modified code snippet.
     You should only modify the part of the code snippet requested in the query and nothing else.
  * Include all the pieces of code necessary to answer the user query.
  * Include atleast a few lines of the surrounding code.
  * Explain if the code semantics remains intact or not and reason about it.
     You *should* only discuss about the **functional equivalence** and nothing
     else.
  * Add line numbers to the original code and modified code.
     Make sure that the line numbers of the original code and modified code
     are the consistent. The line numbers should be adjacent to the code as
     1: <code_statement>
     2: <code_statement>
     ...

    ### Example #1:

    I have retrieved the code for `foo.cc`. There is a use of uninitialized value at
    line 106. Either `i` or `j` is uninitialized.

    ```c++
      // foo.cc
      106|   some_fun(i, j);
    ```

    I need to look at how `i` and `j` are defined and if there are any subsequent
    initializations.

    ### Example #2:

    I have retrieved the code for `foo.cc`. I need to look at bar() function if
    it is relevant and retrieve the code for that.

    ```c++
      // foo.cc
      106|   a = b.bar();
    ```

    I now have to look at the class methods of object b and retrieve the code for bar().
    I should also look at where b is defined.

  * You **must** retrieve the body of functions from "absl::" library using code_search tool for
    those functions that are present in one code snippet and absent in another code snippet.
    You must add this code snippet from the library call to "<Original Code> ... </Original Code>" or
      "<Modified Code> ... </Modified Code>"
  * You must look at lines before and after the annotated part of code and add it to "<Original Code> ... </Original Code>" or
      "<Modified Code> ... </Modified Code>" as required.
  * If there is a function call common to both code snippets, you must add it to Common Code section.
  * You must retrieve every piece of code and put them in "<Original Code> ... </Original Code>",
      "<Modified Code> ... </Modified Code>" or  "<Common Code> ... </Common Code>".
  * If you do not do this, you will be heavily penalized.

  Additional Output Guidelines:
    Your final output should be as follows.
    You should output the relevant code snippets from original and modified code,
    including all the library functions that are required for analysis.
    **DO NOT ADD ANY EXTRA TEXT AS COMMENTS**. You will be penalized if you
    do not use the following format.
      <Original Code>
      </Original Code>

      <Modified Code>
      </Modified Code>

      <Common Code>
      </Common Code>

      <Explanation>
      </Explanation>
\end{lstlisting}

\FloatBarrier
\begin{lstlisting}[caption={System Instruction for Formalizer to generate the predicates for the given code snippets and an explanation of equivalence using the generated predicates.}, label={lst:system-inst-2},language={}]
You are an expert software developer. You are now given two code snippets
-- code 1 and code 2.

* The code snippets are very similar except for minor and local modifications.
* You are asked to create a use-def-control information flow for both code snippets.
* Use the following predicates *as is*
  * def(x : str, src_file : str, line_number : int, code_num: str) : Variable `x` is defined at
    `line_number` in `src_file`. `x` could be a variable 'v' or member property like 'v.y'.
    Any change or update to `x` needs to be captured using this predicate.
    `code_num` indicates whether this predicate is for Code1 or Code2. You should
    only use "Code1" for code 1 and "Code2" for code 2.
    Some function calls on variables can have side effects which update the variable state.
    You should capture this information.
    If a function call foo(x) updates `x` or properties of `x`, you should capture this information.
    Do not use made-up variable names in the predicates or values of the variables.
  * use(x : str, src_file : str, line_number : int, code_num: str) : Variable `x` is used at
    `line_number` in `src_file`. `x` could be a variable 'v' or member property like 'v.y'.
    For every use predicate, there should be a corresponding def predicate for the same variable.
    Do not use made-up variable names in the predicates or values of the variables.
    `code_num` indicates whether this predicate is for Code1 or Code2. You should
    only use "Code1" for code 1 and "Code2" for code 2.
  * flow(x : str, src_file : str, src_line : int, y, dst_file: str, dst_line: int, code_num: str) :
    The value of `x` defined at
    `src_line` in `src_file` flows to the `y` at `dst_line` in `dst_file`.
    For every `x`, there should be a def(x, _) or a flow(_,_, x, _) in your use-def information flow.
    A Flow is usually a result of
    (i) definition reaching a use for the same variable name,
    (ii) an argument being passed at a call site (may or may not have same variable name),
    (iii) assignments (x = y indicates flow from y to x) (may or may not have same variable name),
        even if y and x have the same variable name, you **must** capture the flow.
    (iv) return values (may or may not have same variable name), etc.
    `code_num` indicates whether this predicate is for Code1 or Code2. You should
    only use "Code1" for code 1 and "Code2" for code 2.
  * exit(line : num, code_num: str): `line` is where the main function returns,
    You must generate the exit() predicate for every return in the "main()" function and no other function.
    This predicate is only added for the main relevant code snippet and not for code snippets added to support this code.
    `code_num` indicates whether this predicate is for Code1 or Code2. You should
    only use "Code1" for code 1 and "Code2" for code 2.
  * entry(func : str, line : num, code_num: str): `line` is where the current code or function begins or the provided code snippet starts.
    If the `func` is a "function", then the `line` is where the function definition starts.
    If the current function is not clear, then the `line` is 0.
    `code_num` indicates whether this predicate is for Code1 or Code2. You should
    only use "Code1" for code 1 and "Code2" for code 2.
  * controldep(x : str, src_file : str, src_line : int, cond: str, branch: bool, dst_file: str, dst_line: int, code_num: str) :
    The value of `x` defined at `src_line` in `src_file` *control-depends* on the expression `cond` with choice `branch`
    at `dst_line` in `dst_file`. The `cond` could be "Entry:<function_name>" at the beginning of a function
    or the expression controlling `x` at `src_line`. `branch` depends on "true"/"false" branch.
    The branch should be output as a string -- "true"/"false".
    If the cond is the "Entry", then `branch` will always be true.
    For every def(`x`), there should be a controldep(`x`,) in your output.
    A controldep is usually a result of conditionals (if/else, while) statements.

    The controldep predicate is also generated for every conditional. The variable `x`
    used in the condition at `src_line` in `src_file` *control-depends* on the expression `cond` with choice `branch`
    at `dst_line` in `dst_file`. You must create such predicates for every conditional statement.


    `code_num` indicates whether this predicate is for Code1 or Code2. You should
    only use "Code1" for code 1 and "Code2" for code 2.

    Literature:
    A control dependency directly impacts data-flow analysis, as the set of possible values
    (definitions) that can reach a particular use of a variable is determined
    by the control-flow paths leading to that use. Essentially, control dependencies
    create the framework of paths along which data is tracked and analyzed.

  * defWithExpr(x : str, src_file : str, line_number : int, code_num: str) :
    Variable `x` is defined at `line_number` in `src_file` with an expression.
    This predicate is only used when the assignment to `x` is **not** a constant
    or simple variable assignment. This could imply that the right hand side of the
    assignment is a complex expression.
    `code_num` indicates whether this predicate is for Code1 or Code2. You should
    only use "Code1" for code 1 and "Code2" for code 2.

    Whenever defWithExpr is used, the def predicate should be used as well.

  * isConstantValue(x: str, code_num: str) :
    Variable `x` (a number/enum etc.) is a constant value.
    `code_num` indicates whether this predicate is for Code1 or Code2. You should
    only use "Code1" for code 1 and "Code2" for code 2.

  * unaryFun(function_symbol: str, x: str, src_file: str, line1: int, code_num: str)
    This predicate is generated whenever a defWithExpr or condWithExpr is created.
    This predicate represents an expression or library function with a unary operator
    `function_symbol` applied to `x`.
    Examples include: malloc(a), !v, type(a) etc.
    `code_num` indicates whether this predicate is for Code1 or Code2. You should
    only use "Code1" for code 1 and "Code2" for code 2.
    This predicate is generated only in cases of a complex expression or a function
    that is not present in the code snippet.

  * binaryFun(function_symbol: str, x: str, y: str, src_file: str, line1: int, code_num: str)
    This predicate is generated whenever a defWithExpr or condWithExpr is created.
    This predicate represents an expression or library function with a binary operator
    `function_symbol` applied to `x` and `y`.
    Examples include: a + b, a * b, min(c, d) etc.
    `code_num` indicates whether this predicate is for Code1 or Code2. You should
    only use "Code1" for code 1 and "Code2" for code 2.
    This predicate is generated only in cases of a complex expression or a function
    that is not present in the code snippet.

  * outputVar(x: str, file: str, line: int, code_num: str)
    Variable `x` is an important variable defined or used in the "main" program
    of code_num (Code/Code2) that is necessary to answer the equivalence query.
    You **must** generate this predicate for every variable `x` in the "main" program.
    This predicate is generated at `line` in `file` at the return of the "main" program.
    `code_num` indicates whether this predicate is for Code1 or Code2. You should
    only use "Code1" for code 1 and "Code2" for code 2.

  * varMap(x: str, x_file: str, line1: int, "Code1" : str, y: str, y_file: str, line2: int, "Code2" : str):
    This predicate is generated **only** when there are two variables `x` defined at `line1` in `x_file` in Code1
    and `y` defined at `line2` in `y_file` in Code2 and they are equivalent except for textual differences, i.e,
    for the same intent and same computation, `x` was defined in Code1 and `y` was defined in Code2.

  * exitMap(line1: int, "Code1", str, line2: int, "Code2": str): This predicate
    is generated to create a mapping between the exit(line1) of Code1 and exit(line2)
    of Code2, i.e., which exit (or return from main) in Code1 is similar to the
    exit (or return from main) in Code2. You must create this for every return point
    in the main function.

  * entryMap(line1: int, "Code1", str, line2: int, "Code2": str): This predicate
    is generated to create a mapping between the entry(line1) of Code1 and entry(line2)
    of Code2, i.e., which entry in Code1 is similar to the entry in Code2.


    If you do not follow the signature of the predicates as mentioned, you will
    be penalized. You must represent strings within "" in the predicates.

    You should *only* use variables in the predicates. DO NOT treat function calls as variables.

    You must capture every flow in the code snippet.

    ** Some important points to ensure:
    - Every def(x, line) of variable x predicate should have a corresponding flow(_, line, x, line)
    from some variable (same or different) to x at that line OR it should have a corresponding
    defWithExpr(x, line) predicate and unary/binary predicates depending on the RHS.
    - For every function call in the code snippet, if the function body is present, you **must**
    analyse the function and generate the required predicates (that will be useful in answering equivalence query).
    - You must generate the exit() predicate for every return in the "main()" function and no other function.
    - At every return point in the main() function, you must generate use() predicates for all the **important** variables that you have
      used in your explanation of equivalence and there must be a flow predicate from the previous definition to
      this use for those variables.
    - You must generate the entry() predicate for every "main()" function.
    - If a variable is defined outside the scope of the provided snippet, you should
      assume it is defined elsewhere and add a def() at line where the main function starts,
      it is the same line where entry() for "main" function is generated.
    - For every variable defined in and before the "main" program,
      and at every return point of the "main" program (where exit(l) is generated),
      you must add a use() and flow() predicates for all
      those variables. The flow() predicate should be from the previous def to the current use.
      If there are multiple flows for a variable, you must capture all of them.
      The dst_line should match the line number of the current use and in the generated
      exit predicate.
    - The outputVar() predicate should be generated for every "important" variable
      that effects the state of the final program and which is important in answering
      the equivalence query.
    - The use() and flow() predicates should be generated for every variable with outputVar()
      predicate.


    Example 1: Consider a code snippet as follows:
    ```
    // s.cpp
    1. int x = 1;
    2. if (val1) {
    3.   x = 2;
    4. }
    5. if (val2) {
    6.   x = 3;
    7. }
    8. int y = x;
    9. y = y;
    ```

    The predicates would be:
    ```
    entry(0)
    def("val1", "s.cpp", 0).
    def("val2", "s.cpp", 0).
    def("x", "s.cpp", 1).
    controldep("x", "s.cpp", 1, "Entry:main", "true", "s.cpp", 0)
    use("val1", "s.cpp", 2).
    flow("val1", "s.cpp", 0, "val1", "s.cpp", 2).
    def("x", "s.cpp", 3).
    controldep("x", "s.cpp", 3, "val1", "true", "s.cpp", 2).
    use("val2", "s.cpp", 5).
    flow("val2", "s.cpp", 0, "val2", "s.cpp", 5).
    controldep("x", "s.cpp", 6, "val2", "true", "s.cpp", 5).
    def("x", "s.cpp", 6).
    flow("x", "s.cpp", 1, "x", "s.cpp", 8).
    flow("x", "s.cpp", 3, "x", "s.cpp", 8).
    flow("x", "s.cpp", 5, "x", "s.cpp", 8).
    def("y", "s.cpp", 8).
    flow("x", "s.cpp", 8, "y", "s.cpp", 8).
    use("y", "s.cpp", 9).
    flow("y", "s.cpp", 8, "y", "s.cpp", 9).
    def("y", "s.cpp", 9).
    flow("y", "s.cpp", 9, "y", "s.cpp", 9).
    exit(10)
    use("x", "s.cpp", 10).
    flow("x", "s.cpp", 1, "x", "s.cpp", 10).
    flow("x", "s.cpp", 3, "x", "s.cpp", 10).
    flow("x", "s.cpp", 5, "x", "s.cpp", 10).
    use("y", "s.cpp", 10).
    flow("y", "s.cpp", 9, "y", "s.cpp", 10).
    ```

    Example 2: Consider a code snippet as follows:
    ```
    // s.cpp
    1. int x = 1;
    2. bool c = x == 1;
    3. if (c) {
    4.   return ;
    5. }
    6. return ;
    ```

    The predicates would be:
    ```
    entry(0)
    def("1", "s.cpp", 0).
    isConstantValue("1").
    def("x", "s.cpp", 1).
    controldep("x", "s.cpp", 1, "Entry:main", "true", "s.cpp", 0).
    use("1", "s.cpp", 1).
    flow("1", "s.cpp", 0, "1", "s.cpp", 1).
    flow("1", "s.cpp", 0, "x", "s.cpp", 1).
    def("c", "s.cpp", 2).
    controldep("c", "s.cpp", 2, "Entry:main", "true", "s.cpp", 0).
    use("x", "s.cpp", 2).
    flow("x", "s.cpp", 1, "x", "s.cpp", 2).
    flow("1", "s.cpp", 0, "1", "s.cpp", 2).
    defWithExpr("c", "s.cpp", 2).
    binaryFun("==", "x", "1", "s.cpp", 2).
    use("c", "s.cpp", 3).
    flow("c", "s.cpp", 2, "c", "s.cpp", 3).
    exit(4);
    use("c", "s.cpp", 5).
    flow("c", "s.cpp", 2, "c", "s.cpp", 5).
    use("x", "s.cpp", 5).
    flow("x", "s.cpp", 1, "x", "s.cpp", 5).
    flow("1", "s.cpp", 0, "1", "s.cpp", 5).
    exit(6);
    use("x", "s.cpp", 6).
    flow("x", "s.cpp", 1, "x", "s.cpp", 6).
    use("1", "s.cpp", 6).
    flow("1", "s.cpp", 0, "1", "s.cpp", 6).
    use("c", "s.cpp", 6).
    flow("c", "s.cpp", 2, "c", "s.cpp", 6).
    ```

    Example 3: Consider two code snippets as follows:
    ```
    Code1
    // s.cpp
    1. int x = 1;
    2. int y = x;
    ```
    ```
    Code2
    // s.cpp
    1. int tx = 1;
    2. int y = tx;
    ```

    The predicates corresponding to varMap (all predicates are not shown here) would be:
    ```
    <Common Predicates>
    varMap("x", "s.cpp", 1, "Code1", "tx", 1, "Code2").
    </Common Predicates>
    ```

    Example 4: Consider the code snippet as follows:
    ```
    // s.cpp
    1. int* x = malloc(a);
    2. int y = b + c;
    3. int z = d + e;
    4. int l = b * c;
    ```

    The predicates corresponding to unaryFun and binaryFun (all predicates are not shown here) would be:
    ```
    unaryFun("malloc", "x", "s.cpp", 1, "Code1").
    binaryFun("+", "b", "c", "s.cpp", 2, "Code1").
    binaryFun("+", "d", "e", "s.cpp", 3, "Code1").
    binaryFun("*", "b", "c", "s.cpp", 4, "Code1").
    ```

    Example 5: Consider two code snippets as follows:
    ```
    Code1
    // s.cpp
    1. int a = x;
    2. a = b;
    3. return ;
    ```
    ```
    Code2
    // s.cpp
    1. int y = x;
    2. int a = y;
    3. a = b;
    4. return ;
    ```

    The predicates corresponding to varMap (all predicates are not shown here) would be:
    ```
    <Code1 Predicates>
    outputVar("a", "s.cpp", 3, "Code1").
    </Code1 Predicates>

    <Code2 Predicates>
    outputVar("a", "s.cpp", 4, "Code1").
    </Code2 Predicates>
    ```

* You **should** only use the variable names in the provided code snippets.
* Consider all constants are defined at the start of program at line: 0.
* You **must** used the same file names for both code snippets.
* You **must** end all the predicates with a full stop '.'.
* If you do not know the file name, add "main.cpp" as the file name.
* If a function call is a library function that is not relevant to the question,
  use it as a blacbox and create a formal use-def predicates for the arguments
  of the function call or any global state changes due to the function call
  from the documentation for the function.
* You must do deeper analysis for a function call irrespective of the varMap predicate.
* Finally, you should explain and reason if the code snippets are equivalent or not.
* Your explanation should solely be based on the predicates you generated.
* You cannot use any other information beyond the generated predicates to explain your reasoning.
  You *must* refer to a predicate -- use/def/flow/controldep/exit for every statement made in your explanation.
* If the code snippets are equivalent, your explanation **must** show this through a series of "flow" and "controldep" predicates.
  If they are not equivalent, your explanation **must** show this difference through a series of "flow" and "controldep" predicates.
  You can also use other predicates, particularly "defWithExpr" and "condWithExpr" as needed.
* Give verbose explanations, reasoning about every statement in your explanation only in terms
  of the generated predicates. If you have to repeat yourself, you **must**.

<Code1 Predicates>
</Code1 Predicates>

<Code2 Predicates>
</Code2 Predicates>

<Common Predicates>
</Common Predicates>

<Explanation>
</Explanation>
\end{lstlisting}

\subsection{Explanation with Predicates Examples}
\label{Se:equiv-explain-artifact}
We show an example of the explanation of equivalence (or not) using predicates generated by the Formalizer.~\figrefs{equiv-example-1-code1}{equiv-example-1-code2} shows two code snippets with differ only in two self-assignments. The Formalizer generates the code predicates for the code snippets (not shown in this paper) such as $\tt def(...), flow(...)$ and an explanation of program equivalence using the generated predicates as shown in~\figref{equiv-example-1-explanation}.

\begin{figure}[t!]
    \centering
\begin{subfigure}[t]{0.42\linewidth}
\begin{minipage}[t]{\linewidth} 
    \begin{lstlisting}
// main.cpp
...
189| void main() {
190|  wrk->y_accum += wrk->y_add;
191|  wrk->dst += wrk->dst_stride;
192|  wrk->dst_y += 1;
193|  return ;
194| }
...
    \end{lstlisting}
\end{minipage}%
\caption{Sample code snippet 1}
\label{Fi:equiv-example-1-code1}
\end{subfigure}
\hfill 
\begin{subfigure}[t]{0.55\linewidth}
\begin{minipage}[t]{\linewidth} 
    \begin{lstlisting}
// main.cpp
...
189| void main() {
190|  wrk->y_accum += wrk->y_add;
191|  wrk->dst += wrk->dst_stride;
192|  wrk->num_channels = wrk->num_channels;
193|  wrk->dst_width = wrk->dst_width;
194|  wrk->dst_y += 1;
195|  return ;
196| }
...
    \end{lstlisting}
\caption{Sample code snippet 2}
\label{Fi:equiv-example-1-code2}
\end{minipage}
\end{subfigure}

\hfill

\begin{subfigure}[t]{0.9\linewidth}
\begin{minipage}[t]{\linewidth} 
    \begin{lstlisting}[language={}]
The two code snippets are semantically equivalent. The modifications in Code 2 are self-assignments which do not alter the programs state.

In both code snippets, the program starts at line 189, as indicated by `entry("main", 189, "Code1")` and `entry("main", 189, "Code2")`. The variables `wrk->y_accum`, `wrk->y_add`, `wrk->dst`, `wrk->dst_stride`, and `wrk->dst_y` are defined at the beginning of the execution for both snippets.
...
For `wrk->y_accum`:
- In Code 1, `def("wrk->y_accum", "main", 190, "Code1")` is a result of the expression `wrk->y_accum += wrk->y_add`. The final value of `wrk->y_accum` at `exit(193, "Code1")` is given by `flow("wrk->y_accum", "main", 190, "wrk->y_accum", "main", 193, "Code1")`.
- In Code 2, `def("wrk->y_accum", "main", 190, "Code2")` is a result of the same expression. The final value of `wrk->y_accum` at `exit(195, "Code2")` is given by `flow("wrk->y_accum", "main", 190, "wrk->y_accum", "main", 195, "Code2")`.

For `wrk->dst`:
- In Code 1, `def("wrk->dst", "main", 191, "Code1")` results from `wrk->dst += wrk->dst_stride`. The final value at `exit(193, "Code1")` is given by `flow("wrk->dst", "main", 191, "wrk->dst", "main", 193, "Code1")`.
- In Code 2, `def("wrk->dst", "main", 191, "Code2")` results from the same expression. The final value at `exit(195, "Code2")` is given by `flow("wrk->dst", "main", 191, "wrk->dst", "main", 195, "Code2")`.
...
The additional lines in Code 2, `wrk->num_channels = wrk->num_channels;` and `wrk->dst_width = wrk->dst_width;`, are self-assignments. This is shown by the predicates `flow("wrk->num_channels", "main", 192, "wrk->num_channels", "main", 192, "Code2")` and `flow("wrk->dst_width", "main", 193, "wrk->dst_width", "main", 193, "Code2")`. These operations do not change the values of `wrk->num_channels` and `wrk->dst_width`, and therefore have no impact on the final state of the program.

Since the data flows for all watch variables are preserved and the additional statements are no-ops, the two code snippets are equivalent.
    \end{lstlisting}
\end{minipage}%
\caption{Sample code snippet 1}
\label{Fi:equiv-example-1-explanation}
\end{subfigure}
\caption{The explanation of equivalence generated by Formalizer using the provided predicate signatures.}
\label{Fi:equiv-example-1-output}
\end{figure}

\end{document}